\documentclass{aa}  

\usepackage{amsmath}
\usepackage{physics}
\usepackage{graphicx}
\usepackage[labelfont=bf]{caption}
\usepackage{subfig}
\usepackage{rotating}
\usepackage{txfonts}
\usepackage{makecell}
\usepackage{multirow}
\usepackage{hyperref}
\hypersetup{
    colorlinks,
    linkcolor={blue!80!black},
    citecolor={blue!80!black},
    urlcolor={blue!80!black}
}
\usepackage{natbib}
\bibpunct{(}{)}{;}{a}{}{,}
\usepackage{nicefrac}
\usepackage{dblfloatfix}
\usepackage{boldline}
\usepackage[dvipsnames]{xcolor}
\usepackage{nicefrac}

\newcommand{\Mp}{M$_{P}$}

\newcommand{\Msun}{\textrm{M}$_\mathrm{\odot}$}

\newcommand{\Mj}{$M_J$}

\begin{document}

   \title{The quest for Magrathea planets}

   \subtitle{II. Orbital stability of exoplanets formed around double white dwarfs.}

   \author{Arianna Nigioni 
          \inst{\ref{unige}}
          \and
          Diego Turrini \inst{\ref{oato},\ref{icsc}}
          \and
          Camilla Danielski \inst{\ref{unival},\ref{oaa}}
          \and
          Danae Polychroni \inst{\ref{oato},\ref{icsc}}
          \and
          John E. Chambers \inst{\ref{carnegie}}
        }

           \institute{Observatoire astronomique de l’Université de Genève, Chemin Pegasi 51, CH-1290 Versoix, Switzerland \label{unige} \\ \email{arianna.nigioni@unige.ch}
            \and
            INAF - Osservatorio Astrofisico di Torino, via Osservatorio 20, 10025, Pino Torinese\label{oato}
            \and
            ICSC – National Research Centre for High Performance Computing, Big Data and Quantum Computing, Via Magnanelli 2, Casalecchio di Reno, 40033, Italy\label{icsc}
            \and
            Departament d'Astronomia i Astrof\'{i}sica, Universitat de Val\`{e}ncia, C. Dr. Moliner 50, 46100, Burjassot, Spain \label{unival}
            \and
            INAF - Osservatorio Astrofisico di Arcetri, Largo E. Fermi 5, 50125, Firenze, Italy \label{oaa}
            \and
            Carnegie Institution for Science, Earth and Planets Laboratory, 2541 Broad Branch Road NW, Washington DC 20015,USA\label{carnegie}
             }

   \date{}

  \abstract
   {
   Planetary formation might occur at different stages of the stellar evolution. In particular, theoretical studies have been focusing on addressing whether formation can happen around compact binaries that evolved beyond the Main Sequence. Formation of second-generation planets has been tested in circumbinary discs formed by the ejection of stellar material from binaries composed by either a main sequence star and a white dwarf, or a double white dwarf (DWD). In the latter case, formation appears to be common, creating both sub-Neptunian, Neptunian and giant planets which can migrate within 1 au of the central binary. Nevertheless, studies on the orbital stability of such systems have yet to be undertaken.
   }
   {
   In this work we investigate whether planetary systems formed around compact DWDs, in both non-resonant and resonant configurations, can be dynamically stable over a timescale of a few million years. 
   }
   {
   We perform N-body simulations of circumbinary multi-planetary systems initially hosting two, three, four or five planets by employing an hybrid symplectic integrator made specifically for circumbinary systems. We record the catastrophic events that planetary systems experience and employ a variety of metrics, e.g., orbital spacing, variation of the center of mass and Normalized Angular Momentum Deficit (NAMD), to explore the outcomes of their long term evolution. 
   Furthermore, we evaluate the potential for detecting these systems in their final configurations with the Laser Interferometer Space Antenna (LISA) mission by measuring the overall gravitational-wave frequency shift amplitude induced by their planets.
   }
   {
   Our results show that planets orbiting DWDs can be stable over the studied timescales. While planetary systems starting with two-planets are more likely to survive unaltered, planetary systems with three, four or five planets, experience catastrophic events that cause them to lose some of their original planets. At the end of their phases of dynamical instability the five-planet population is completely disrupted and most of the systems end up hosting only two surviving planets. This increases the number of two-planet systems by 122\% with respect to their initial abundance and creates a single-planet population amounting to 7\% of the totality of systems. Additionally, the four-planet population decreases by 56.1\% and the three-planet population by 22.5\%. Finally, 7.7\% of systems end up being disrupted and they initially hosted more than two planets. Finally, we find that the majority of the systems that end up with a single planet, are potential candidates for LISA. Concerning multi-planet systems, a handful of systems could be detected. Finally, we provide a formula to estimate the gravitational wave frequency shift amplitude for multi/planet systems orbiting DWDs.
   }
   {
   Throughout our analysis we highlight the importance of both characterizing the systems' orbits and estimating their NAMD in order to distinguish between the different dynamical scenarios presented above. Ultimately, second-generation systems may represent crucial targets for the LISA mission, as they reside within its range of observability.
   }

    
   \titlerunning{Orbital stability of Magrathea exoplanets}
   \maketitle

\section{Introduction}
Present observational biases only allow for a small number of circumbinary planets (CBPs or P-type planets) to be detected, yet the occurrence rate of this population is theorized to be comparable or larger than the occurrence rate of planets orbiting single Main Sequence stars \citep{Armstrong2014}. So far, we have 16 fully confirmed CBPs in 13 binary systems. While the majority has been discovered with the transit method thanks to the Kepler and TESS missions (e.g., \citealt{Doyle2011,Triaud2022,Martin2018,Kostov2020,Kostov2021,Socia2020}), 3 planets (TOI-1338/BEBOP-1 c, BEBOP-3 b, \citealt{Standing2023,Baycroft2025}) have been detected in the through Radial Velocity by the BEBOP (Binaries Escorted By Orbiting Planets) survey \citep{Martin2019}).

Studies by \cite{kostov} and \cite{Columba2023} showed that CBPs present a larger probability of survival throughout the evolution of their stellar hosts compared to their single-host siblings, and that a percentage (23-32\%) of planets around close binaries composed of Main Sequence stars (whose separation is less than 10 au) can survive the evolution of both binary components until the  white dwarf (WD)  stage \citep{Columba2023}.
Among the detected systems, planets have been found orbiting binaries at different evolutionary stages: from main sequence binaries (e.g., \citealt{Doyle2011,Kostov2020,Kostov2021}) to binaries composed by one WDs and one low-mass star (e.g., \citealt{Beuermann2010,Potter2011}); exoplanets around double white dwarfs (DWDs) are yet to be observed \citep{dani2019,Katz2022}.  

An interesting aspect concerning systems containing evolved stars is that,
throughout the evolution of the inner binary, there is the possibility of forming a disc from the material ejected out of the common-envelope (CE) phase \citep{vanwinckel,Cava2021}. In particular, \citet{kashiandsoker} showed that 1-10\% of the ejected envelope remains bound to the binary and falls back forming a circumbinary disc due to angular momentum conservation. 
This portion of CE ejected material that forms the circumbinary disc, also called ``second-generation disc'', is gas-rich and contains dust \citep{kluska}.
As a result, these second-generation discs can have mass that are comparable to first-generation pre-Main Sequence ones, providing favorable conditions to form second-generation massive exoplanets \citep{Perets2010}.
The presence of a giant planet could also explain the observed depletion of refractory elements in the circumbinary disc, compared to the volatile abundances \citep{kluska}. 
More recently \citet{Ledda2023} explored the possibility of planetary formation around DWDs and showed that in this type of environment giant planets can form with a planetary mass of at least \Mp\ $\sim 0.27$\Mj. These planets are 
identified as ``second-generation'' ones because they form within second-generation discs\footnote{It is important to point out that a more appropriate denomination of discs and planets forming in the last stage of the binary evolution would be ``third-generation''.}.

This body of work argues that planets around DWDs, also identified as $Magrathea$ planets \citep{TamaniniDanielski2019}, should exist either because (i) they are first-generation exoplanets which survived the binary evolution \citep{kostov,Columba2023}, or because (ii) they have formed afterwards from a second-generation disc surrounding a DWD \citep{Ledda2023}.

Studies of the stability of circumbinary planets around binaries highlight how the gravitational field of the binary creates a network of Mean Motion Resonances (MMR)\footnote{An orbital MMR between two planets occurs when their orbital periods are nearly commensurate, i.e. they can be expressed as a ratio of two small integers \citep{lissauer}, that are responsible for the presence of chaotic dynamical regions that surround the inner binary (see \citealt{marzari19} for a review on the topic).
}
In these regions, existing planets would be ejected and no new planets would form. Therefore, the final distribution of planets depends on the gravitational influence that the binary companion has on them, but also on dynamical processes such as migration, tidal interactions and/or planet-planet scattering \citep{marzari19}.

Concerning P-type systems, the unstable region is located closer to the inner stars, where perturbations are stronger \citep{marzari19}. A widely used analytical expression for the limit of this region was found by \citet{WiegertHolman} in the framework of the restricted three-body problem. Their work was then extended by \cite{Quarles2018} and \cite{Adelbert2023}, who widened the range of the sampled parameter space. The recent work of \citet{Georgakarakos2024} addressed hierarchical triple systems, providing a different framework to study the stability of circumbinary planets.

The stability limit, identified by $a_{\text{crit}}$, represents the semi-major axis beyond which planetary orbits are stable with respect to the stellar perturbations. In the case of multi-planet systems, however, the perturbations among planets can excite their orbits and destabilize the system, leading to catastrophic events like planetary collisions or ejections. 

Stability and dynamical excitation studies of exoplanetary systems showed that a large fraction of them already experienced or will experience a chaotic evolution \citep{LimbachTurner15,LaskarPetit17,ZinziTurrini17,turrini2020,turriniB,gajdos2023}. Observations of the circumbinary extrasolar planets found by Kepler show that their orbits are, in the majority of cases, eccentric and suggest that the secular evolution of the system was affected by planet-planet scattering \citep{gong}. When many close encounters occur, the orbits of the planets can become highly eccentric and inclined \citep{gong} and trigger more violent phases of dynamical instability and chaotic evolution. As a result, even if second-generation planetary systems could form, it is currently unclear whether they would survive long enough to be observable or whether they should be expected to become unstable and be destroyed.

This work builds on the results of \cite{Ledda2023} and focuses on studying the stability of giant exoplanets orbiting the center of mass of various DWD systems. Over the past few years,  several studies have demonstrated the potential of the ESA Laser Interferometer Space Antenna (LISA) to detect such $Magrathea$ exoplanets across the Milky Way  \citep{seto2008,TamaniniDanielski2019,dani2019,Katz2022} and even in the Large Magellanic Clouds \citep{DanielskiTamanini2020}. 
Our work is carried out within the framework of the LISA science preparation activities \citep{LISAcallpaper,AmaroSeoane2022:WP}, specifically contributing to the development of the planetary detection science case.
Moreover, the results presented here are also relevant for the scientific planning of other space-borne gravitational-wave observatories, such as Taiji \citep{kang2021:Taiji} and TianQin \citep{Huang2020:TianQin,Hu2025:TianQin}.

The paper is organized as follows: 
in Sec. \ref{sec:methods} we describe the 
methodology of the study, the N-body algorithm used in the simulations and the metrics employed to analyze their output.
In Sec. \ref{sec:results} we present our results by identifying four Populations (A, B, C and D), and by analyzing their differences in terms of the parameter space defined by the metrics. Finally, we discuss our results in Sec. \ref{discussion} and draw conclusions in Sec. \ref{sec:conclusions}.

\section{Methods}\label{sec:methods}
In this section we describe the symplectic algorithm for integrating the equations of motion of P-type planetary systems, the setup of the N-body simulations and the metrics we use to analyze their output.

\subsection{Symplectic mapping of P-type binary systems}\label{sec:symplclosebin}

The symplectic mapping method, first introduced by \citet{WisdomHolman}, is based on solving an approximated version of Hamilton's equations of motion that describe the evolution of an N-body system. In its simplest implementation, the symplectic mapping splits the Hamiltonian $H$ into two components: one describing the pure Keplerian evolution of the planetary bodies around the star, while the other accounts for the perturbations due to the interactions between the planetary bodies \citep[e.g.,][]{Chamb99}. As long as the perturbation term is smaller than the Keplerian one, the symplectic mapping method is characterized by the desirable properties of being fast while guaranteeing the long-term energy conservation of the simulated system.

Symplectic integrators are not suitable for simulating systems where planets can undergo collisions or close encounters. The solution given by \cite{Chamb99} to this limitation is to numerically integrate the term that becomes dominant during the close encounter so that the integration accuracy does not get compromised. Following \citet{Chamb99}, the integration for those time-steps when a close encounter takes place is done with the Bulirsh-Stoer integrator implemented in the \textsc{Mercury} package. This class of modified algorithms is known as hybrid symplectic integrators and proves to be an optimal choice for planet formation simulations. While the symplectic mapping and the hybrid algorithm here described have been designed for single star systems, \cite{Chamb02} introduced a new set of coordinates and equations that are adapted to study binary systems while also allowing to handle close encounters. For a detailed description of this algorithm and its related equations, we refer to \cite{Chamb02}. In the following we we will focus on equation not explicitly described in \cite{Chamb02}, but that are integral part of the algorithm.\\

\noindent The positions $\vec{X}$ and conjugate momenta $\vec{P}$ are defined by Eqs. (16) and (17) of \cite{Chamb02}. The momenta $\vec{P}$ are related to pseudo-velocities $\vec{V}$ rather than the real velocities $\dot{\vec{X}}$, and these quantities are connected by the equations
\begin{equation}
    \begin{split}
        \vec{P}_A &=m_{\text{tot}}\vec{V}_A\\
        \vec{P}_B &=m_{B}\vec{V}_B\\
        \vec{P}_i &=m_{i}\vec{V}_i\\
    \end{split}
\end{equation}
\begin{equation}\label{vel&pseudo}
    \begin{split}
        \vec{V}_A &=\dot{\vec{X}}_A\\
        \vec{V}_B &=\dfrac{m_A}{m_{\text{bin}}}\dot{\vec{X}}_B\\
        \vec{V}_i &=\dot{\vec{X}}_i-\dfrac{\sum_j m_j\dot{\vec{X}}_j}{m_{\text{tot}}}\\
        \dot{\vec{X}}_i &= \vec{V}_i+\dfrac{\sum_j m_j\vec{V}_j}{m_{\text{bin}}}
    \end{split}
\end{equation}

The inverse transformation from P-type coordinates $(\vec{X},\vec{V})$ to inertial coordinates $(\vec{x},\vec{v})$, are given by \cite{Verrier}:
\begin{equation}\label{closetoinertialX}
    \begin{split}
        \vec{x}_A &= \vec{X}_A-\dfrac{m_B}{m_{\text{bin}}}\vec{X}_B-\dfrac{\sum_j m_j\vec{X}_j}{m_{\text{tot}}}\\
        \vec{x}_i &=\vec{X}_A+\vec{X}_i-\dfrac{\sum_j m_j\vec{X}_j}{m_{\text{tot}}}\\
        \vec{x}_B &= \vec{X}_A-\dfrac{m_A}{m_{\text{bin}}}\vec{X}_B-\dfrac{\sum_j m_j\vec{X}_j}{m_{\text{tot}}}
    \end{split}
\end{equation}
\begin{equation}\label{closetoinertialV}
    \begin{split}
        \vec{v}_A &= \vec{V}_A-\dfrac{m_B}{m_A}\vec{V}_B-\dfrac{\sum_j m_j\vec{V}_j}{m_{\text{bin}}}\\
        \vec{v}_i &=\vec{V}_i+\vec{V}_A\\
        \vec{v}_B &= \vec{V}_A+\vec{V}_B-\dfrac{\sum_j m_j\vec{V}_j}{m_{\text{bin}}}
    \end{split}
\end{equation}  

\noindent One important point in implementing the algorithm is that while the effects of the interaction and the jump Hamiltonians, defined by Eqs. (19) and (22) of \cite{Chamb02}, are treated in terms of the coordinates from Eqs. (16) and (17) of \cite{Chamb02}, the evolution of the Keplerian Hamiltonian should be computed using the velocities $\vec{\dot{X}}$ from Eq. \ref{vel&pseudo} in place of the pseudo-velocities. This can be achieved by converting back and forth the pseudo-velocities using Eq. \ref{vel&pseudo} but numerical tests showed that, when these frequent conversions are applied to the binary companion, they significantly degrade the accuracy of the symplectic algorithm. A computationally efficient solution that allows to avoid this issue consists in using the coordinates from Eqs. (16) and (17) of \cite{Chamb02} while adopting the modified central mass $m_A^2/m_{\text{bin}}$ and the modified time-step $(m_{\text{bin}}/m_A)\cdot\tau/2N_{\text{bin}}$ for the binary companion (J. E. Chambers, priv. comm.). The integration of the planetary Keplerian motion is not affected by this issue, therefore we use the coordinates $\vec{X}$ and velocities $\vec{\dot{X}}$ with gravitational central mass $m_{\text{bin}}$ and time-step $\tau$.

\subsection{Simulations setup}\label{sec:simusetup}
In our simulations we analyze the stability of orbits around the stellar binaries reported in Table \ref{tab:DWDparams}, where the systems marked with the $*$ are drawn from the LISA DWD Population \citep{Korol2019}. These specific binaries were selected because they were used by \cite{Ledda2023}, whose simulated planet formation outcomes provide the input parameters for our study. $M_1$ and $M_2$ are the stellar masses, $P_{\text{bin}}$ is the period of the binary orbit, $a_{\text{bin}}$ is the binary separation and $a_{\text{crit}}$ is the critical semi-major axis introduced in Section \ref{subsec:criticalSMA} and defined by equation \eqref{acrit}. The binary eccentricity is set to zero in all simulated systems (see Table 1 of \citealt{Ledda2023} for more details) and the inclinations are set to zero for simplicity. In each simulation, we randomly draw the argument of pericenter and mean anomaly between 0$^\circ$ and 360$^\circ$, while the longitude of the ascending node is drawn randomly between 0$^\circ$ and 180$^\circ$.
We define the planetary system priors as described in the following subsections. 

\subsubsection{Critical semi-major axis}\label{subsec:criticalSMA}
The gravitational influence of a binary system is responsible for regions of instability for a planet orbiting in a P-type architecture. In the context of the restricted three body problem, an analytical expression of the critical semi-major axis $a_{\text{crit}}$ that delimits the unstable area was formulated by \citet{WiegertHolman} with their equation (3). They considered massless particles on initially circular and prograde orbits in the binary plane of motion. The binary mass ratio $\mu=M_2/(M_1+M_2)$ was taken between 0.1 and 0.9, the binary eccentricity $e_{\rm bin}$ ranged from 0 to 0.7 and their simulations lasted $10^4$ binary periods. The results of \cite{WiegertHolman} were later extended by \cite{Quarles2018}, who expanded the parameter space by considering binary mass ratios down to 0.01, binary eccentricities up to 0.8 and increasing the integration time to $10^5$ binary periods. Furthermore, the planetary bodies were not considered massless and their mean anomaly was also sampled. The expression found by \cite{Quarles2018} for the critical semi-major axis is (Fit 2 from their Table 2)

\begin{equation}\label{acrit}
      \begin{split}
          a_{\text{crit}}= & ~a_{\text{bin}} \cdot[\qty(0.93\pm0.02)+(2.67\pm0.08)~e_{\rm bin}~ + \\
          & + \qty(-0.25\pm0.06)~e_{\rm bin}^2 + \qty(3.72\pm0.06)~\mu^{1/3} + \\ & + \qty(2.25\pm0.12)~\mu^{1/3} e_{\rm bin} +\qty(-2.72\pm0.05)~\mu^{2/3} + \\
          & + \qty(-4.17\pm0.15)~\mu^{2/3} e_{\rm bin}^2]~
      \end{split}
\end{equation}

\noindent where $a_{\text{bin}}$ is the orbital separation between the two stars. Orbits with a semi-major axis $a$ larger than $a_{\text{crit}}$ are stable with respect to the stellar perturbations. 

The later work by \cite{Adelbert2023} simulated binaries with the same mass ratios as \cite{WiegertHolman} while considering planetary eccentricities up to 0.9. Their equation (4) gives their best fit for the critical semi-major axis (which is defined $r_\text{c,peri}$ by the authors). However, they did not sample different initial mean or true anomalies since the planet was always initialized at the same position with respect to the binary star.

In their recent paper, \cite{Georgakarakos2024} introduce a new framework for studying the stability of P-type planets in hierarchical triple systems, significantly expanding the sampled parameter space. Unlike previous works, they considered both coplanar and non-coplanar configurations, including cases such as perpendicular orbits, and simulating both prograde and retrograde orbits. Additionally, they investigated the dependence of the stability boundary from the planet mass, considering planets with masses up to 1\% of the mass of the binary. They also investigated a wide range of orbital eccentricities, from nearly circular to highly elliptical (up to e=0.9), and extended the simulations to a duration of $10^6$ planetary orbital periods. When addressing the stability of their simulated systems they consider three different regimes:  i) stable motion, ii) mixed stable-unstable motion and iii) unstable motion. As a result, they define two critical semi-major axes: one corresponding to the transition between regimes (i) and (ii) and the other corresponding to the transition between regimes (ii) and (iii). For further details and clarifications we refer to Section 2.2. of \cite{Georgakarakos2024}. Their empirical fits for the two said stability limits are given by their equation (10). The parameter that had the greatest effect on the stability of the systems proved to be the planetary eccentricity. Specifically, higher planetary eccentricity values yield higher values for the two critical borders.

Since the parameter that produces the largest difference between the formulations of \cite{Quarles2018} and \cite{Georgakarakos2024} is the planetary eccentricity and given that the initial eccentricity values of the planetary orbits in our simulations are quite small ($e<0.01$), we adopted the simplest formulation of eq. \eqref{acrit} to set our initial conditions.

\begin{table*}[t]
    \caption{Stellar masses, periods, binary separations and critical semi-major axis of the three analyzed DWDs.}
    \label{tab:DWDparams}
    \centering
    \begin{tabular}{l|c|c|c|c|c}
        \hline
        \hline
         system & $M_1\ [M_\odot]$ & $M_2\ [M_\odot]$ & $P_{\text{bin}}$ [h] & $a_{\text{bin}}$ [au] & $a_{\text{crit}}$ [au]  \\
         \hline
         DWD$_2$ & 0.31 & 0.21 & 0.43 & 1.078$\times10^{-3}$ & 2.36$\times10^{-3}$ \\
         DWD$_3^*$ & 0.75 & 0.26 & 1.51 & 3.106$\times10^{-3}$ & 6.82$\times10^{-3}$ \\
         DWD$_4^*$ & 0.31 & 0.25 & 1.71 & 2.773$\times10^{-3}$ & 6.06$\times10^{-3}$ \\
         \hline
    \end{tabular}
    \tablefoot{The binary eccentricity is zero for all the systems (see \citealt{Ledda2023}, their Table 1 for more details). The inclinations are set to zero for simplicity. The systems marked with the $*$ belong to the LISA DWD Population presented in \citet{Korol2019}. $a_{\text{crit}}$ is calculated with equation \eqref{acrit}.}
\end{table*}

\subsubsection{Initial conditions for the planetary systems}\label{subsec:initialconditions}

In initializing the planetary systems of our simulations we consider both non-resonant and resonant architectures, with the non-resonant ones representing the nominal set-up.

\subsubsection*{Nominal set-up}
The planets that formed in the work of \cite{Ledda2023} have masses $M_p$ that range between [0.4,0.5] \Mj, [2.5,15] \Mj\ and [0.12,1.2] \Mj, for DWD$_2$, DWD$^*_3$ and DWD$^*_4$ respectively. Given that our focus is  exploring the survivability of systems hosting second-generation giant planets from \cite{Ledda2023}, we extract the planet masses from uniform distributions in these ranges. For each binary system, we run 500 simulations with a time-step of 2.5\% of the orbital period of the innermost planet. The initial multiplicity $N_s$ is drawn between 2 and 4 and the percentage of systems with $N_s=2,3,4$ is given in Table \ref{tabpianetiiniziali}. During the evolution, if the orbital distance of a planet grows larger than 100 au from the center of mass of the binary, that planet is considered as "scattered/ejected". This means that the planet will be either ejected over long time-scales or it will survive on a far-out orbit that we are not able to observe. For this reason we remove the planet from the simulation. Furthermore, when a planet approaches the central binary within a distance equivalent to the binary separation, determined using the planet's interpolated trajectory during the time-step, we do not resolve the physical collision, as it is beyond the scope of this study. Instead, these cases are reclassified as unstable due to a "close approach to the binary", and the planet is subsequently removed from the system. In the DWD$^*_3$ and DWD$^*_4$ study cases, the integration lasts for 10$^7$ years and the orbital parameters are recorded every 10$^4$ years. In the DWD$_2$ case, integration lasts $3\times10^6$ years with the same output interval as the previous cases. The difference is motivated by the fact that DWD$_2$ is more compact (see Tab. \ref{tab:DWDparams}), and therefore the innermost planet can reside on a closer orbit compared to the other binary configurations. Depending on the actual period of the planet, $N_\mathrm{bin}$ can grow to a very large number and mimic the effect of a much larger multiplicity of the system, which substantially increases the computational cost and leads to a slower integration. To overcome this fact, the system is studied for a comparable number of time-steps, hence the shorter duration. These timescales remains fully appropriate for our purposes, as recent dynamical studies \citep[e.g.,][]{Izidoro2017} indicate that major dynamical instabilities typically occur early in planetary system evolution, shortly after disc dispersal.

\begin{table}[h]
    \caption{\label{tabpianetiiniziali}Percentage of systems with initially 2, 3 and 4 planets.}
    \centering
    \begin{tabular}{l|c|c|c}
        \hline
        \hline
         & $N_s=2$ & $N_s=3$ & $N_s=4$\\
         \hline
         DWD$_2$ - nominal& 24\% & 56\% & 20\%\\
         DWD$^*_3$ - nominal & 16\% & 62\% & 22\%\\
          DWD$^*_3$ - resonant & 40\% & 52\% & 8\%\\
         DWD$^*_4$ - nominal & 20\% & 54\% & 26\%\\
         \hline                  
    \end{tabular}
\end{table}

The planetary separation $a$ of the first planet is always drawn between $a_{\text{crit}}$ and 1 au, in order to have at least one planet within 1 au from the binary. The other planets are positioned between $a_{\text{crit}}$ and 
$a_{\text{L}}=(G\cdot P^2(M_{\text{bin}}+M_p)/4\pi^2)^{1/3}$ where $M_{\text{bin}}$ is the total mass of the binary and the period $P=8$ years accounts for a possible extension of the LISA nominal mission.We note that $a_{\text{crit}}$ lies well beyond the radius where tides and General Relativity effects are being exerted. When initializing the semi-major axis of each planet, we imposed the separation from all other planets to be greater than the minimum mutual separation for the stability of each pair ($r_{\text{H}_{i,i+1}}$, \citealt{Chamb96,Obertas2017}). Given $a_{i,i+1}$, the planetary semi-major axes for planet $i$ and planet $i+1$, this separation is expressed in units of mutual Hill radii as \citep{Obertas2017}.
\begin{equation}
    \dfrac{\abs{a_{i}-a_{i+1}}}{\Delta_{\text{C}}\cdot r_{\text{H}_{i,i+1}}}>1
\end{equation}
\begin{equation}
\begin{split}
    &\text{where}\quad \Delta_{\text{C}}=\dfrac{2\sqrt{3}}{1+2\sqrt{3}\cdot X}\quad\text{and}\quad X=\dfrac{1}{2}\qty(\dfrac{M_{p,i}+M_{p,i+1}}{3M_{\text{bin}}})^{1/3}
\end{split}
\end{equation}
\noindent The initial inclination among planetary orbital planes is randomly initialized within 1$^\circ$ - 3$^\circ$ range, while the eccentricity is randomly drawn between $e$ = 0 - 0.01, justified by the fact that newly formed 
planetary systems are expected to be dynamically cold \citep{Turrini19,turriniA}. The adopted ranges are equivalent to assuming the equipartition of dynamical excitation \citep{LaskarPetit17,turrini2020}.We simulate 50 architectures that differ in terms of $M_p$, $N_s$, $a$, $e$, $i$, sampled as described above. For each architecture we simulate 10 realizations adopting different argument of pericenter, mean anomaly and longitude of the ascending node. The first two phase angles are drawn randomly between 0$^\circ$ and 360$^\circ$, while the longitude of the ascending node is drawn randomly between 0$^\circ$ and 180$^\circ$.
Our simulations do not include Kozai-Lidov effects or any relativistic effect (e.g., production of gravitational waves) that can cause changes in the orbit of the inner binary \citep[e.g.,][]{nel01,LipunovPostnov1988,TutukovYungelson1979}. 

\subsubsection*{Resonant set-up}

In our resonant set-up we simulated two kinds of systems:

\begin{enumerate}
    \item  Systems with initially two, three or four planets, with each adjacent pair in a 2:1 resonances, orbiting the DWD$^*_3$ binary.
    \item[2] Five-planet systems in one of the resonant cases of \cite{Pzryluski2025}, which in turn are based on one of the resonant systems from \cite{Nesvorny2012}. The adopted architecture is a combination of 4:3 and 5:3 resonances. Following \cite{Pzryluski2025} we will label this set-up as NMS resonance. We simulate this systems around the DWD$^*_4$ binary.
\end{enumerate}

In the first case we randomly extract the initial number of planets, their initial mass, eccentricity, inclinations and phase angles as described for the nominal set up. For the initial semi-major axis, we randomly draw the location of the innermost planet between $a_{\text{crit}}$ and 1 au, and then we place the remaining planets by imposing that adjacent pairs are in a 2:1 resonance.

For the second scenario, we simulate the evolution of the planetary system with masses and semi-major axes as reported in Table \ref{NMS_resonance}. They correspond to the set-up of \cite{Pzryluski2025}, where the planet semi-major axes are scaled down by a factor 15. We chose this scaling factor to have all five planets initially within an orbital period of 4 years (i.e., the duration of the nominal LISA mission) and, given that the planet masses fall only in the range of the DWD$^*_4$ system, we employed this specific binary. We then sampled the eccentricity, inclination and phase angle values as in the other cases presented above.

In total, for both resonant scenarios, we simulate 500 systems, i.e. 50 architectures that differ in terms of $M_p$, $N_s$, $a$, $e$, $i$ for the former case and in terms of $e$, $i$ for the latter, each characterized by 10 sets of phase angle values.

As for the nominal DWD$^*_3$ and DWD$^*_4$ binaries, the integration lasts 10$^7$ years and the time-step is set to 2.5\% of the orbital period of the innermost planet. Furthermore, we apply the same conditions of planet removal due to ejection and/or close approach to the binary as described before.

\begin{table}[h]
    \caption{\label{NMS_resonance}Initial planet mass and semi-major axes for the resonant five-planet systems orbiting the DWD$^*_4$ binary.}
    \centering
    \begin{tabular}{l|c|c}
        \hline
        \hline
         Planet & $M_p$ [$M_J$]& $a$ [au]\\
         \hline
         Jupiter & 1.0 & 5.71/15 \\
         Saturn & 0.3 & 7.78/15 \\
         Planet 5 & 0.045 & 10.51/15 \\
         Uranus & 0.045 & 17.62/15 \\
         Neptune & 0.053 & 23.34/15 \\
         \hline                  
    \end{tabular}        
\end{table}

\subsection{Metrics}\label{sec:metrics}

In order to characterize the stability and evolution of each DWD planetary system we used four dimensionless quantities: the Normalized Angular Momentum Deficit (NAMD, \citealt{Chamb01,turrini2020}), the number of bodies in the system ($N$), the fraction of the total mass that is retained in the largest object ($S_m$) and the orbital spacing statistics ($S_s$) from \citet{Chamb01}. 

The NAMD value does not directly indicate whether a planetary system is currently stable or unstable, but rather quantifies its overall dynamical excitation as a consequence of its past evolution, in a normalized form that allows comparison across different architectures. As discussed by \citet{turrini2020,turriniB}, the NAMD value of the Solar System ($1.3\times10^{-3}$) can be considered a boundary between orderly and chaotic evolutionary histories, since the Solar System’s current architecture has been proposed (though not definitively confirmed) to have resulted from a past dynamical instability. Systems with NAMD values higher than that of the Solar System are therefore more likely to have experienced instability phases and catastrophic events, while those with lower values likely underwent a more ordered, stable evolution. The stability analysis by \citet{Rickman2023} further supports the validity of this approach. 

Intuitively, the NAMD value can be interpreted as a dynamical temperature \citep{turriniB}: the bigger the NAMD, the more excited the system is and, adopting a thermodynamic analogy, the resulting dynamical temperature is higher, i.e. the system is dynamically hotter.

The NAMD is defined as: 

\begin{equation}\label{NAMD}
    \text{NAMD}=\dfrac{\sum_k m_k\sqrt{a_k}\qty(1-\sqrt{1-e_k^2}\cos{i_k})}{\sum_k m_k\sqrt{a_k}}
\end{equation}

\noindent where $m_k$, $a_k$ and $e_k$ are the masses, semi-major axis and eccentricities of the planetary bodies in the system. The NAMD provides a uniform measure of dynamical excitation to compare systems with different architectures. We note that the NAMD does not provide information regarding collisions among bodies or their ejection from the system, for such we separately track the occurrences of these events (see Section \ref{sec:results}).

The parameters $S_m$ and $S_s$ \citep{Chamb01} are defined as
\begin{equation}\label{Sm}
    S_m=\dfrac{M_{\rm MP}}{\sum_i^N m_i}
\end{equation}
\begin{equation}\label{Ss}
    S_s=\dfrac{6}{N-1}\qty(\dfrac{a_{\text{max}}-a_{\text{min}}}{a_{\text{max}}+a_{\text{min}}})\qty(\dfrac{3M_{\text{bin}}}{2\bar{m}})^{1/4}\quad\text{where}\quad N>1
\end{equation}

\noindent where $M_{\text{MP}}$ is the mass of the most massive planet in the system, $m_i$ is the mass of each planet $i$,
$\bar{m}$ is the mean mass of the planetary system, and finally $a_{\text{max}}$ and $a_{\text{min}}$ are the semi-major axes of the outermost and innermost planet, respectively.

We computed the normalized ratios $S_m^f/S_m^s$ and $S_s^f/S_s^s$ between the start (i.e., superscript $s$) and at the end (i.e., superscript $f$) of the simulations. The first ratio measures whether the most massive object retains the biggest mass fraction of the system, whereas the $S_s^f/S_s^s$ gives the information of whether the system undergoes an expansion or becomes more compact. Furthermore, with the initial and final number of planets present in each system, $N_s$ and $N_f$ respectively, we compute the number of lost planets in each systems $N_s - N_f$ .\\
Finally, we also compute where the center of mass (CoM) of the planetary system is at the beginning  and the end of the simulation 
\begin{equation}\label{COM}
    \text{CoM}=\dfrac{\sum_i^N m_i a_i}{\sum_i^N m_i}
\end{equation}
Evaluating the difference $\Delta_{\rm CoM}$, we can assess whether the barycenter of the planetary system, without considering the binary, shifts inward or towards the outer regions. This is informative on which planets are removed from the systems: whether outer planets (i.e., $\Delta_{\rm CoM}<0$), or the inner ones (i.e., $\Delta_{\rm CoM}>0$).

\section{Results}
\label{sec:results}
We simulated the evolution of four different Populations, two coming from the nominal non-resonant set-up and two from the resonant one:
\begin{itemize}
    \item Population A: Systems with planets that have initial mass $M_p\leq1.2$\ \Mj\ and that orbit a binary with mass $M_{\text{bin}}< 0.6\ $\Msun. These correspond to the DWD$_2$ and DWD$^*_4$ planetary systems, therefore Population A is made of 1000 simulations, 500 for each binary system. 
    \item Population B: Systems with planets that have initial mass $M_p\geq2.5$\ \Mj\ that orbit a binary with $M_{\text{bin}}\sim1$\ \Msun. These correspond to the 500 DWD$^*_3$ planetary systems. 
    \item Population C: Resonant counterpart of Population B. These correspond to the 500 DWD$^*_3$ planetary systems that are initially in a 2:1 resonance.
    \item Population D: Five-planet systems that are initially in a NMS resonance. These correspond to the 500 DWD$^*_4$ resonant planetary systems.
\end{itemize}

\noindent for which we summarize our results in Tables \ref{tabEventiPopAB}, \ref{tab.ejectedbodAB} and \ref{tabeventiNAMDmagAB}. In the following we will discuss the evolution of these Populations in terms of whether their final state is dynamically excited or not, and whether they underwent catastrophic events resulting in the loss or removal of planets. When discussing catastrophic collisions, we will identify planet-planet collisions with "collisions$_\text{p-p}$". Additionally, when a planet is removed due to a close approach with the binary, we will write "close approach to the binary", while if it is removed because its orbital distance exceeds 100 au then we will write "ejections/scatterings".

\begin{table}
\caption{\label{tabEventiPopAB}Percentage of Population A, B, C and D planetary systems which experienced catastrophic events.}
    \centering
    \begin{tabular}{l||c|c}
        \hline
        \hline
        \multicolumn{3}{c}{Nominal set-up}\\
        \hline
        & \multicolumn{1}{c|}{Pop. A} & \multicolumn{1}{c}{Pop. B}\\
        \hline
         No catastrophic events & 82.5\% & 49\% \\
         Ejections/scatterings only & 8.6\% & 20.6\% \\
         Collisions$_\text{p-p}$ only & 6.1\% & 15.2\% \\
         Close approach to the binary & 0.8\% & 0.2\%\\
         Multiple catastrophic events & 2\% & 15\% \\
        \hline
        \hline
        \multicolumn{3}{c}{Resonant set-up}\\
        \hline
        & \multicolumn{1}{c|}{Pop. C} & \multicolumn{1}{c}{Pop. D}\\
        \hline
         No catastrophic events & 65.6\% & 0\% \\
         Ejections/scatterings only & 11.2\% & 39.2\% \\
         Collisions$_\text{p-p}$ only & 7\% & 4.8\% \\
         Close approach to the binary & 0\% & 0\%\\
         Multiple catastrophic events & 16.2\% & 56\% \\
         \hline
    \end{tabular}
    \tablefoot{Population A multiple catastrophic events (2\%) arise from 1.8\% of ejections/scatterings+collisions$_\text{p-p}$, 0.1\% of ejections/scatterings+close approach to the binary and 0.1\% of ejections/scatterings+collisions$_\text{p-p}$+close approach to the binary; Population B multiple catastrophic events (15\%) arise from: 5.4\% of ejections/scatterings+collisions$_\text{p-p}$, 7.4\% of ejections/scatterings+close approach to the binary, 0.4\% of collisions$_\text{p-p}$+close approach to the binary and 1.8\% of ejections/scatterings+collisions$_\text{p-p}$+close approach to the binary; Population C multiple catastrophic events (16.2\%) arise from 4.6\% of ejections/scatterings+collisions$_\text{p-p}$, 7.8\% of ejections/scatterings+close approach to the binary, 0.2\% of collisions$_\text{p-p}$+close approach to the binary and 3.6\% of ejections/scatterings+collisions$_\text{p-p}$+close approach to the binary; Population D multiple catastrophic events (56\%) arise from 38.4\% of ejections/scatterings+collisions$_\text{p-p}$, 11\% of ejections/scatterings+close approach to the binary and 6.6\% of ejections/scatterings+collisions$_\text{p-p}$+close approach to the binary. 
    }
\end{table}

\begin{table}
    \caption{\label{tab.ejectedbodAB} 
    Percentage of Population A, B, C and D planetary systems that loose 0, 1, 2, 3, 4 or 5 planets.} 
    \centering
    \begin{tabular}{c||c|c|c|c}
        \hline\hline
         $N_s-N_f$ & Pop. A & Pop. B & Pop. C & Pop. D\\
        \hline
         0 & 82.5\% & 49\% & 65.6\% & 0\%\\
         1 & 10.3\% & 26.6\% & 11.4\% & 3.2\%\\
         2 & 5.1\% & 11.6\% & 9.6\% & 12.2\%\\
         3 & 1.9\% & 9\% & 10.2\% & 58\%\\
         4 & 0.2\% & 3.8\% & 3.2\% & 9.4\%\\
         5 & - & - & - & 17.2\%\\
         \hline
    \end{tabular}
\end{table}

\begin{table}
    \caption{\label{tabeventiNAMDmagAB} Percentage of systems that experienced catastrophic events over the subsample of excited Population A, B, C and D systems (i.e., NAMD>1.3$\times10^{-3}$).}
    \centering
    \begin{tabular}{l||c|c}
        \hline
        \hline
        \multicolumn{3}{c}{Nominal set-up}\\
        \hline
        & \multicolumn{1}{c|}{Pop. A} & \multicolumn{1}{c}{Pop. B}\\
        \hline
         No catastrophic events & 9.8\% & 35\% \\
         Ejections/scatterings only & 55.1\% & 28.5\% \\
         Collisions$_\text{p-p}$ only & 19.2\% & 15.8\% \\
         Close approach to the binary & 3.2\% & 0.3\%\\
         Multiple catastrophic events & 12.7\% & 20.4\% \\
        \hline
        \hline
        \multicolumn{3}{c}{Resonant set-up}\\
        \hline
        & \multicolumn{1}{c|}{Pop. C} & \multicolumn{1}{c}{Pop. D}\\
        \hline
         No catastrophic events & 18.6\% & 0\% \\
         Ejections/scatterings only & 26.3\% & 40.1\% \\
         Collisions$_\text{p-p}$ only & 16.3\% & 2.5\% \\
         Close approach to the binary & 0\% & 0\%\\
         Multiple catastrophic events & 38.8\% & 57.4\% \\
         \hline
    \end{tabular}
    \tablefoot{Population A multiple catastrophic events (12.7\%) arise from 11.5\% of ejections/scatterings+collisions$_\text{p-p}$, 0.6\% of ejections/scatterings+close approach to the binary and 0.6\% of ejections/scatterings+collisions$_\text{p-p}$+close approach to the binary; Population B multiple catastrophic events (20.4\%) arises from: 7.2\% of ejections/scatterings+collisions$_\text{p-p}$, 10.2\% of ejections/scatterings+close approach to the binary, 0.6\% of collisions$_\text{p-p}$+close approach to the binary and 2.4\% of ejections/scatterings+collisions$_\text{p-p}$+close approach to the binary; Population C multiple catastrophic events (38.8\%) arise from 11\% of ejections/scatterings+collisions$_\text{p-p}$, 18.7\% of ejections/scatterings+close approach to the binary, 0.5\% of collisions$_\text{p-p}$+close approach to the binary and 8.6\% of ejections/scatterings+collisions$_\text{p-p}$+close approach to the binary; Population D multiple catastrophic events (57.4\%) arise from 39.2\% of ejections/scatterings+collisions$_\text{p-p}$, 11.4\% of ejections/scatterings+close approach to the binary and 6.8\% of ejections/scatterings+collisions$_\text{p-p}$+close approach to the binary.}
\end{table}

\subsection{Population A}\label{subsec:PopulationA}

\begin{figure}
   \centering
   \includegraphics[width=\hsize]{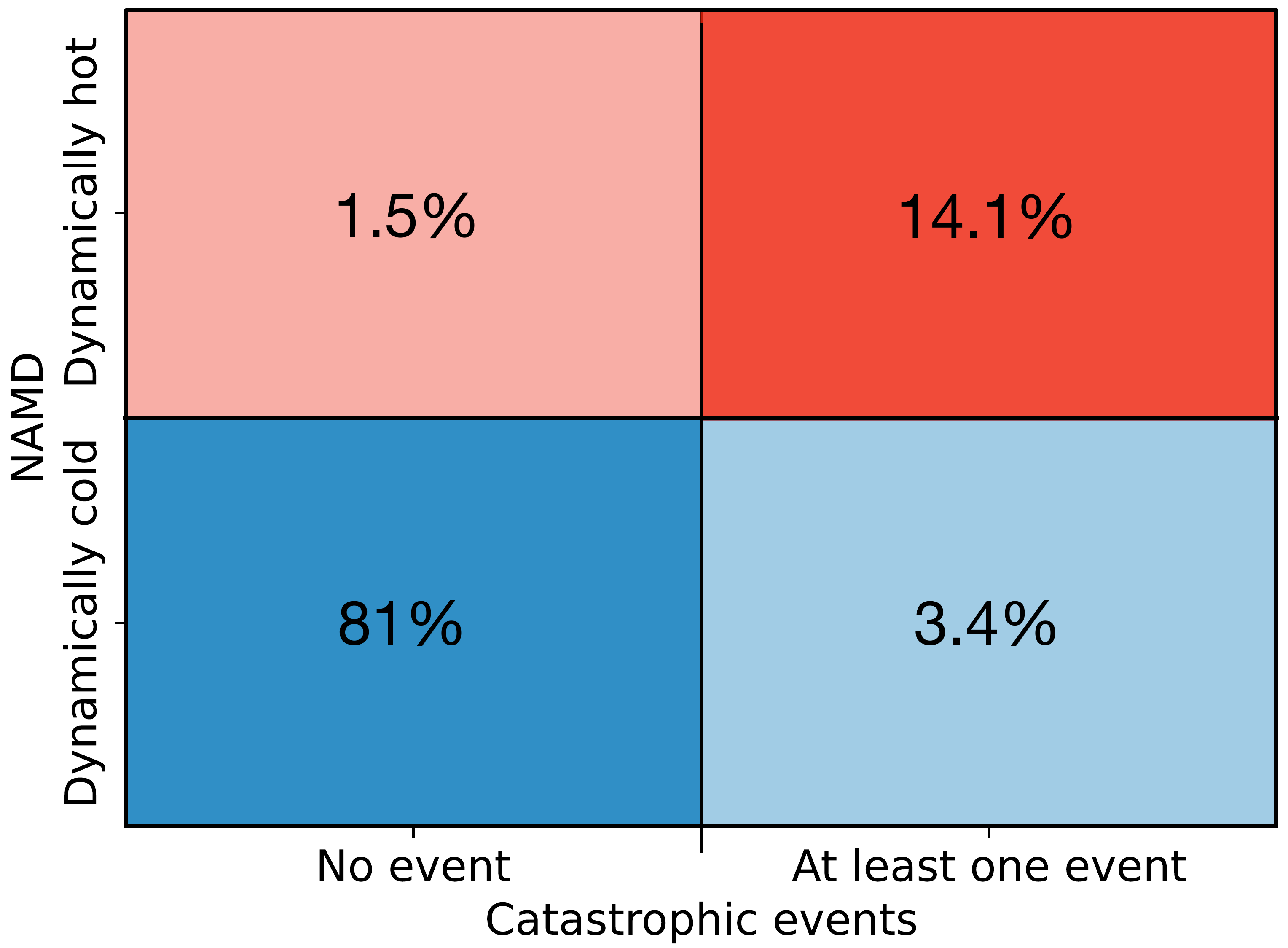}
      \caption{Population A: schematic representation of the percentage of simulated planetary systems (1000 simulations in total for DWD$_2$ and DWD$^*_4$) based on two specific properties: whether they experience or not at least one catastrophic event (ejection/scattering/collision) and whether they are dynamically cold or hot. The number in each box is related to the percentage of systems which have the combination of parameters specified on the x an y axes. The 14.1\% of systems displayed in upper right corner include the 0.2\% of systems that are disrupted (see text for details).}
         \label{heatmapA}
\end{figure}

Population A accounts for the least-massive, non-resonant, planetary systems. Among these systems, over the simulated timescales 84.4\% remain dynamically cold (NAMD < 1.3$\times10^{-3}$) while 15.6\% become dynamically hot (NAMD > 1.3$\times10^{-3}$). In terms of catastrophic events, we find that 82.5\% conserves their initial number of planets (Tab. \ref{tabEventiPopAB} and \ref{tab.ejectedbodAB}) while 17.5\% of systems go through a violent chaotic phase that causes them to experience the loss of planets during their evolution. Only 1.5\% of systems end up dynamically excited without undergoing planet loss (see Fig. \ref{heatmapA} for a schematic representation). Furthermore, 0.2\% of planetary system ends up being completely disrupted, i.e. losing all planets from LISA's observational window.

With the adopted priors, the most common catastrophic event is planetary ejection (or scatter to extremely wide orbits), which takes place alone in 8.6\% of cases, while in combination with other catastrophic events in 2\% of the cases (Tab. \ref{tabEventiPopAB}). 

When we normalize the occurrence rates of catastrophic events to the frequency of dynamically excited systems (i.e., those with NAMD > $1.3\times10^{-3}$, representing 15.6\% of the Population A systems), 55.1\% of the excited planetary systems experiences ejections/scatterings only, while in 12.7\% of cases ejections take place in combination with other catastrophic events (Tab. \ref{tabeventiNAMDmagAB}). Only 9.8\% of the dynamically excited systems do not lose any planet, meaning they cross a phase of instability but do not experience catastrophic events.

\begin{figure*}
    \centering
    \includegraphics[width=0.95\columnwidth]{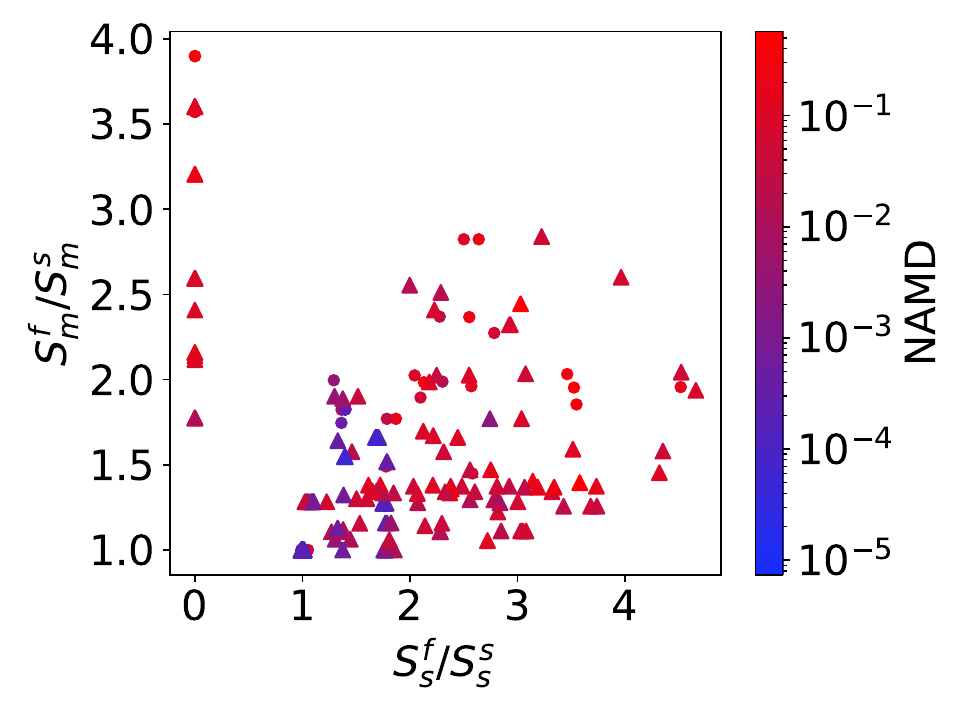}
    \;\;
    \includegraphics[width=0.95\columnwidth]{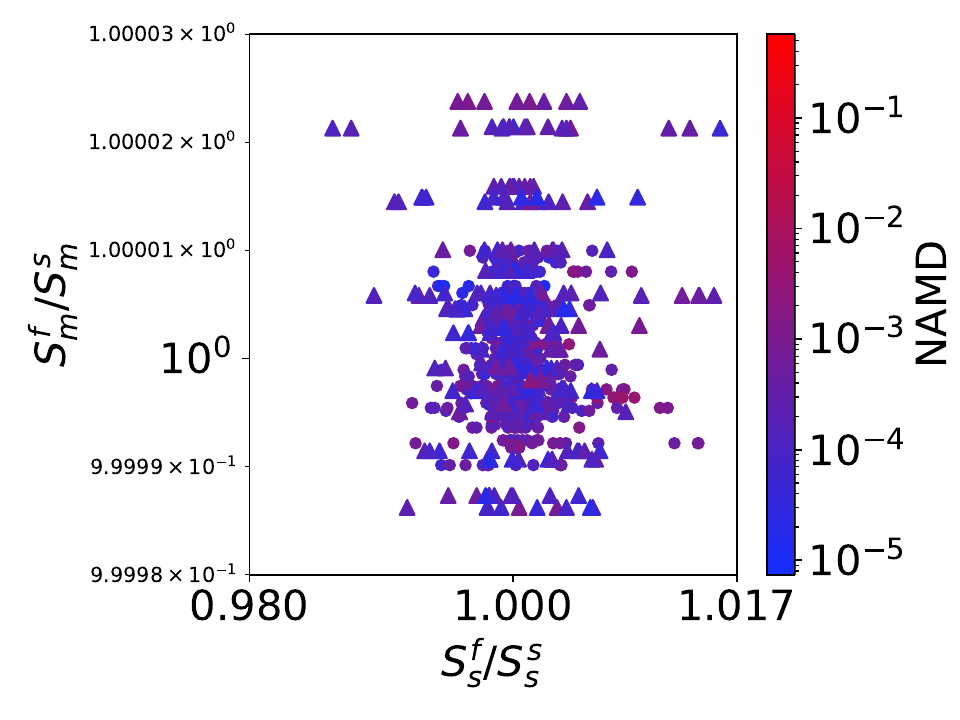}
    \;\;
    \includegraphics[width=0.95\columnwidth]{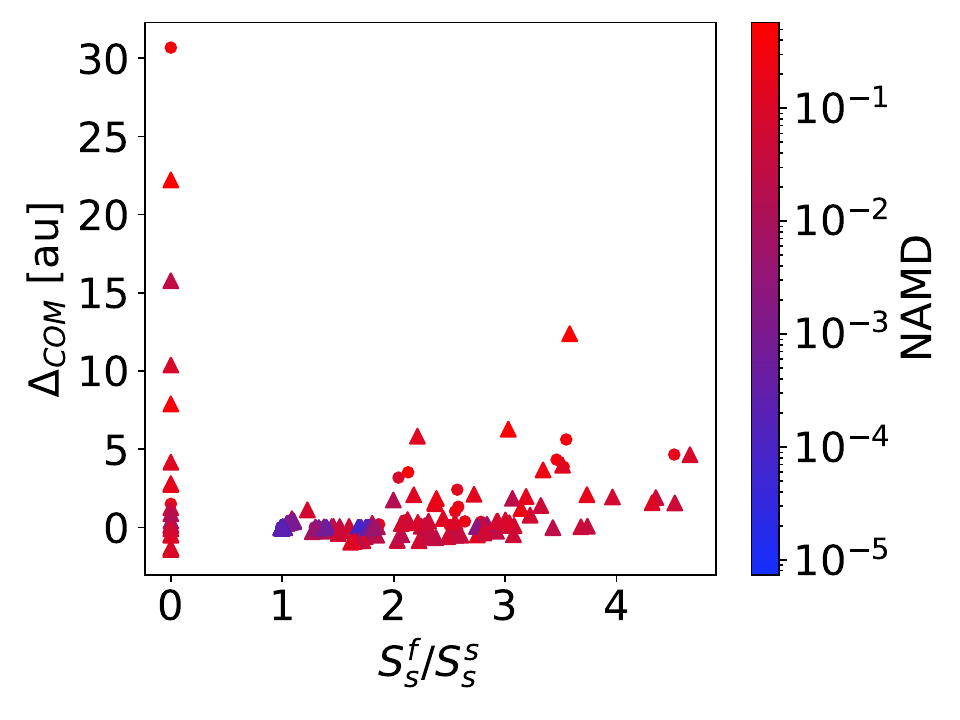}
    \;\;
    \includegraphics[width=0.95\columnwidth]{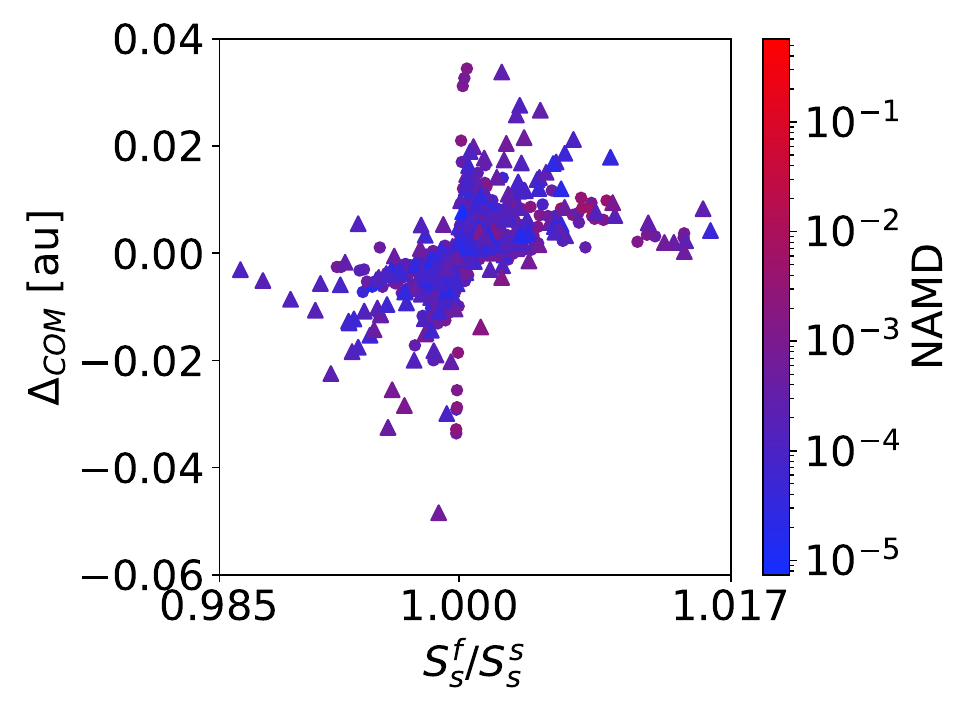}
    \vspace{-0.3cm}
    \caption{Population A: $S_m^f/S_m^s$ as a function of $S_s^f/S_s^s$ (top left panel) with a zoom-in of the region where both ratios are equal or very close to unity (top right panel) and $\Delta_{\rm CoM}$ as a function of $S_s^f/S_s^s$ (bottom left panel) with a zoom-in of the region where $S_s^f/S_s^s$ is equal or very close to unity and $\Delta_{\rm CoM}$ is equal or very close to zero (bottom right panel). DWD$_2$ and DWD$^*_4$ systems are displayed as circles and triangles, respectively. Disrupted systems are not displayed in this figure as it was not possible to compute their NAMD and $\Delta_{\rm CoM}$.}
    \label{SsvsSmCOM_A}
\end{figure*}

When we focus on the metrics $S_m^f/S_m^s$ and $S_s^f/S_s^s$, we see that systems that do not undergo catastrophic events within the simulated timescales are characterized by negligible changes. As can be expected for systems undergoing instabilities, the systems that experience the loss of planets have larger mass concentration in the most massive planet, and generally become less packed (i.e., their orbital spacing increases and $S_s^f/S_s^s>1$) as illustrated by the top left panel in Fig. \ref{SsvsSmCOM_A}. While the loss of planes is expected to reduce the NAMD \citep{Chamb01}, we find that this effect is not strong enough to erase the dynamical signatures of the instability on the dynamical excitation of the planetary systems. This result is consistent with the multiplicity-NAMD anti-correlation observed by \cite{turrini2020} for multi-planet systems around main sequence stars. All systems with $S_s^f/S_s^s < 1$ are either systems with only one surviving planet for which the $S_s^f/S_s^s = 0 $, or stable systems that experienced small variations in their orbital configuration due to secular evolution (top right panel in Fig. \ref{SsvsSmCOM_A}). 

\begin{figure}
\centering
    \includegraphics[width=\columnwidth]{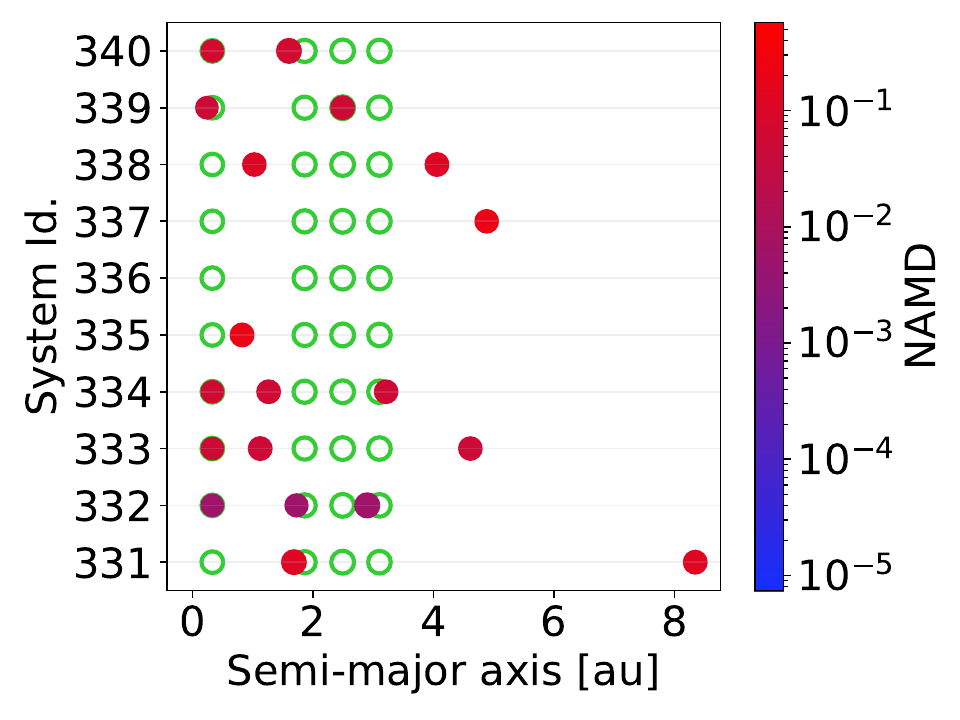}
    \includegraphics[width=\columnwidth]{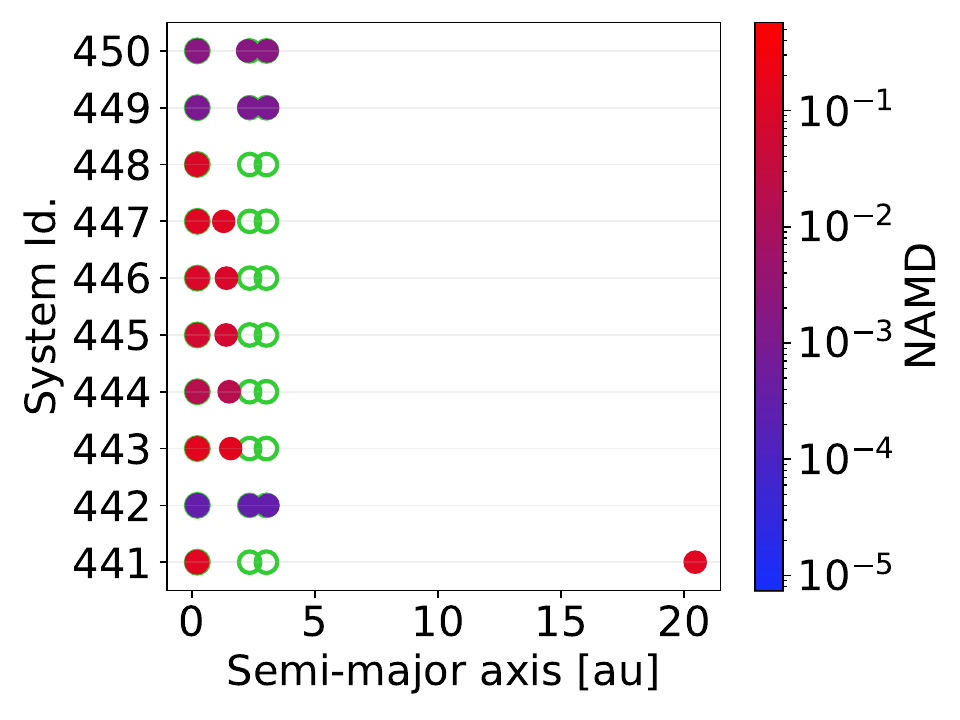}
    \vspace{-0.6cm}
    \caption{DWD$^*_4$ systems (Population A): initial and final $a$ (empty green circles, filled points, respectively). Systems IDs \#331-350 (on the vertical axis, top panel), System IDs \#441-450 (on the vertical axis, bottom panel). Each point represents a planet and its size is proportional to the planetary mass; the color-map gives a measure of the degree of stability. Note that in some systems the green circles overlap with the filled points.}
    \label{posPopA_merged}
\end{figure}

Comparing the $S_s^f/S_s^s$ ratio to the variation of the CoM, $\Delta_{\rm CoM}$, systems with $S_s^f/S_s^s$ close to unity have $\Delta_{\rm CoM}$ close to zero (bottom right panel in Fig. \ref{SsvsSmCOM_A}). This result is expected since systems that are dynamically colder usually do not experience catastrophic events and do not change significantly their orbital separation. Instead, planets that increase their orbital separation during the simulation can experience either a positive shift in the CoM or a negative one. An illustrative example of a system with $S_s^f/S_s^s>1$ and $\Delta_{\rm CoM}>0$, is provided in the top left panel of Figure \ref{posPopA_merged} by system ID \#338 that has initially four planets. The two innermost ones are ejected, while the two surviving ones increase their orbital distance. Alternative scenarios are shown in the top panel (ID \#331) and in the bottom panel (ID \#441), where the outermost planet shifts markedly outward compared to its initial position, consequently causing an outward shift of the CoM. Illustrative examples of systems with $S_s^f/S_s^s>1$ and $\Delta_{\rm CoM}<0$ are shown in the bottom panel of the same Figure \ref{posPopA_merged}, where the systems \#443-447 show that while the outermost planet is ejected, the surviving outer planet moves closer to the binary compared to its original position. In the same panel we also show examples of dynamically cold systems (IDs \#442 and \#449).

\subsection{Population B}\label{subsec:PopulationB}
\begin{figure}
   \centering
   \includegraphics[width=\hsize]{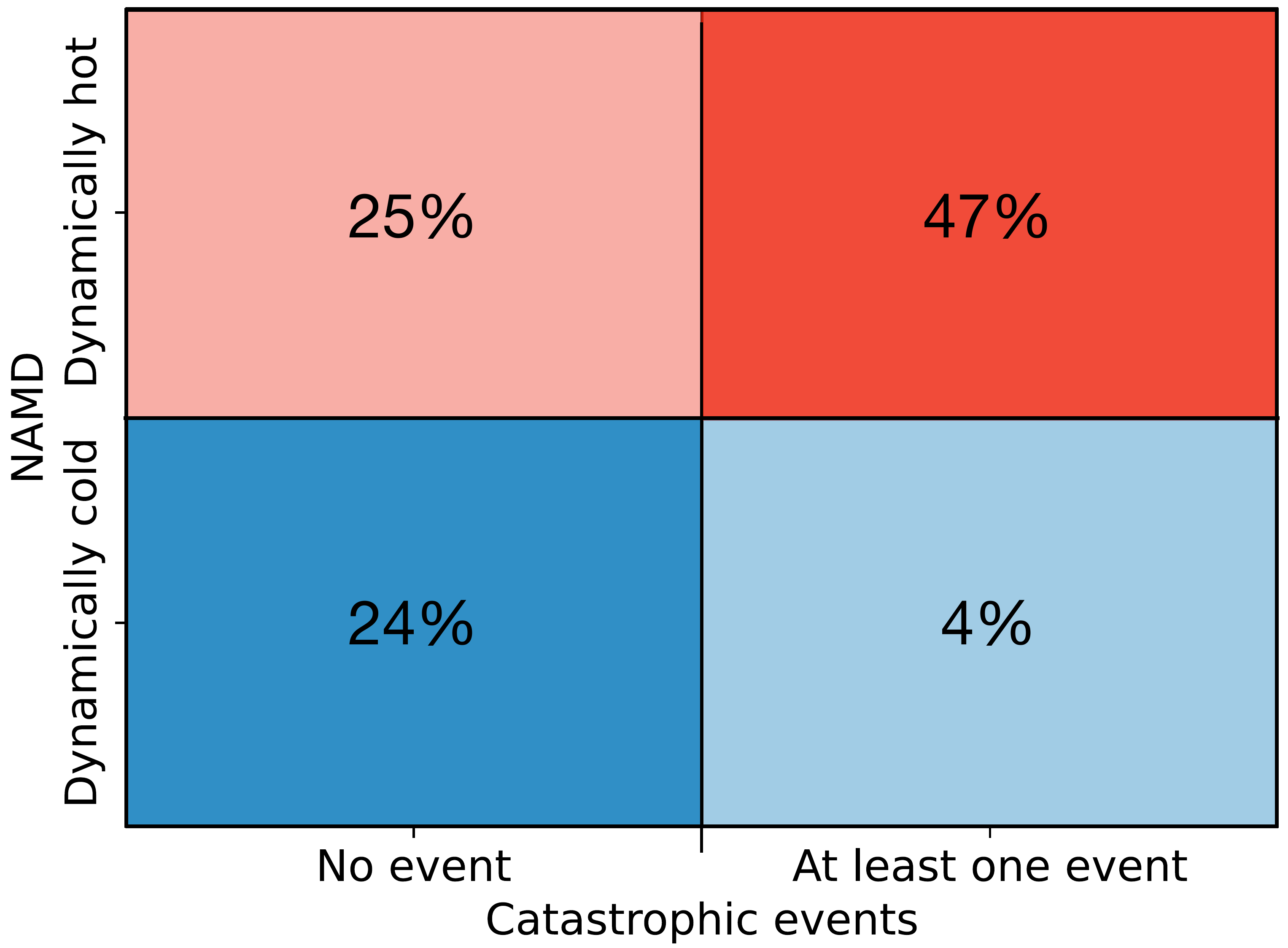}
      \caption{Population B: schematic representation of the percentage of simulated Population B planetary systems (500 simulations) based on two specific properties: whether they experience or not at least one catastrophic event (ejection/scattering/collision) and whether they are dynamically cold or hot. The number in each box is related to the percentage of systems which have the combination of parameters specified on the axes. The 47\% of systems displayed in upper right corner include the 9\% of systems that are disrupted (see text for details).}
         \label{heatmapDWD3}
\end{figure}

When we shift our focus on the more massive, non-resonant, planetary Population B, we find that comparatively fewer planetary configurations remain stable and dynamically cold over the simulated timescales (28\% of the total, see Fig. \ref{heatmapDWD3}). Planetary systems that evolve into dynamically hot architectures represent 72\% of the total, with 47\% of these simulated systems experiencing catastrophic events during their evolutions and 9\% of the simulated systems ending up completely disrupted. As in the case of Population A, we find that a small fraction of systems (4\% in this case) ends up remaining dynamically cold even if experiencing catastrophic events.

For the adopted priors, we find that 49\% of the simulated systems do not lose any planet (Tab. \ref{tabEventiPopAB}) while about 5 out of 10 systems whose planetary multiplicity decreases, experienced the loss of more than one planet (Table \ref{tab.ejectedbodAB}). About 2 out of five systems that experienced catastrophic events are affected by ejections, which in three cases out of five occurs in combination with other catastrophic events.

When we normalize the occurrence rates of catastrophic events to the number of dynamically excited systems in Population B, we find that two out of three of those systems that undergo instabilities experience multiple catastrophic events.

\begin{figure*}
    \centering
    \includegraphics[width=0.95\columnwidth]{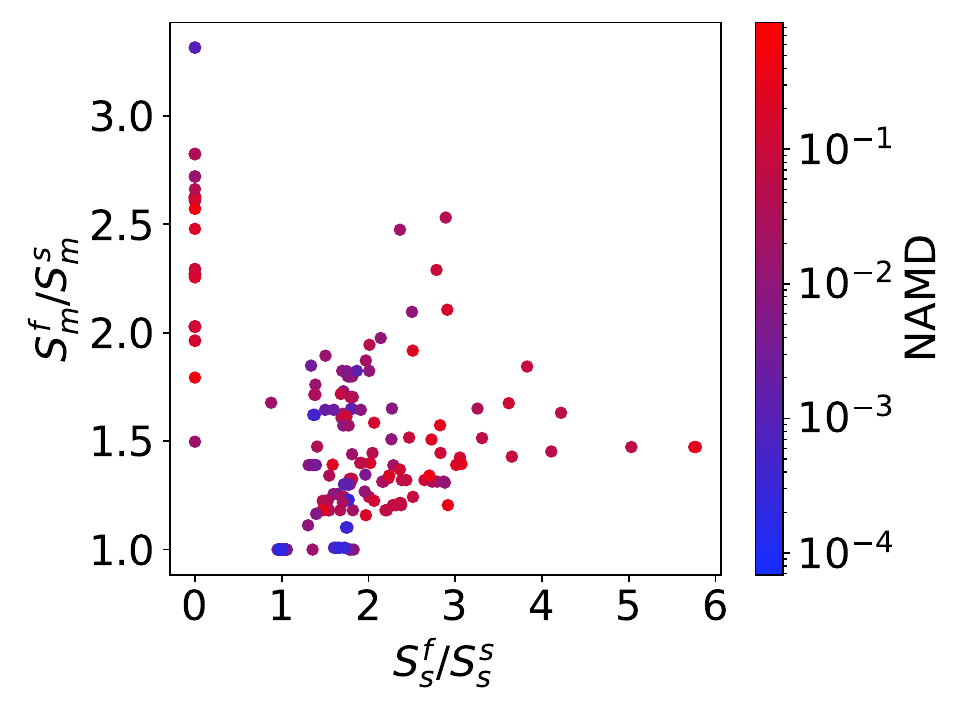}
    \;\;
    \includegraphics[width=0.95\columnwidth]{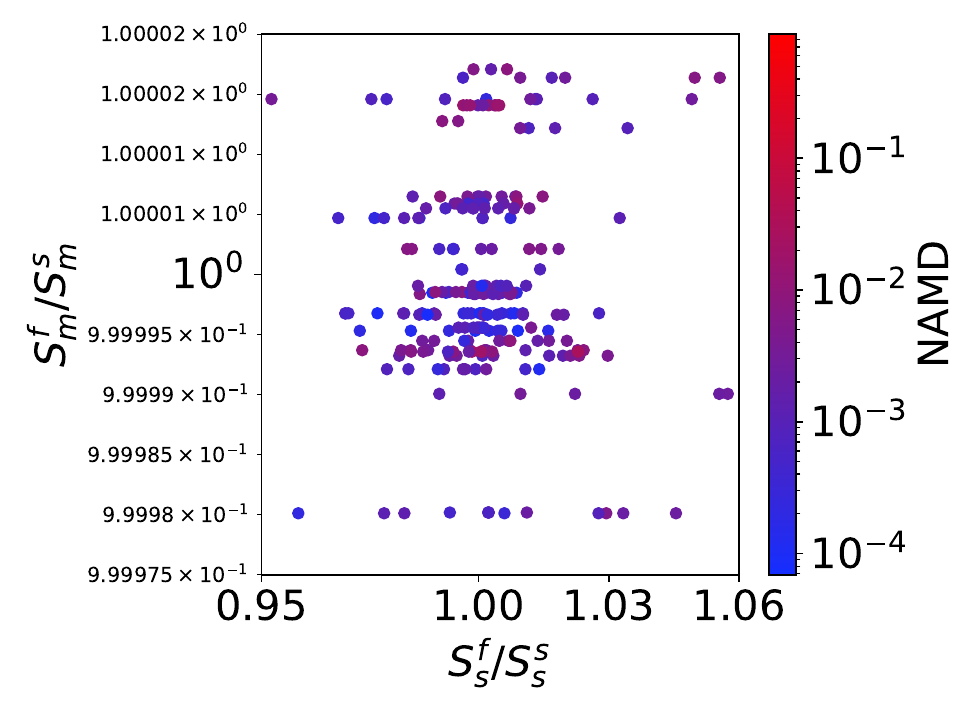}
    \;\;
    \includegraphics[width=0.95\columnwidth]{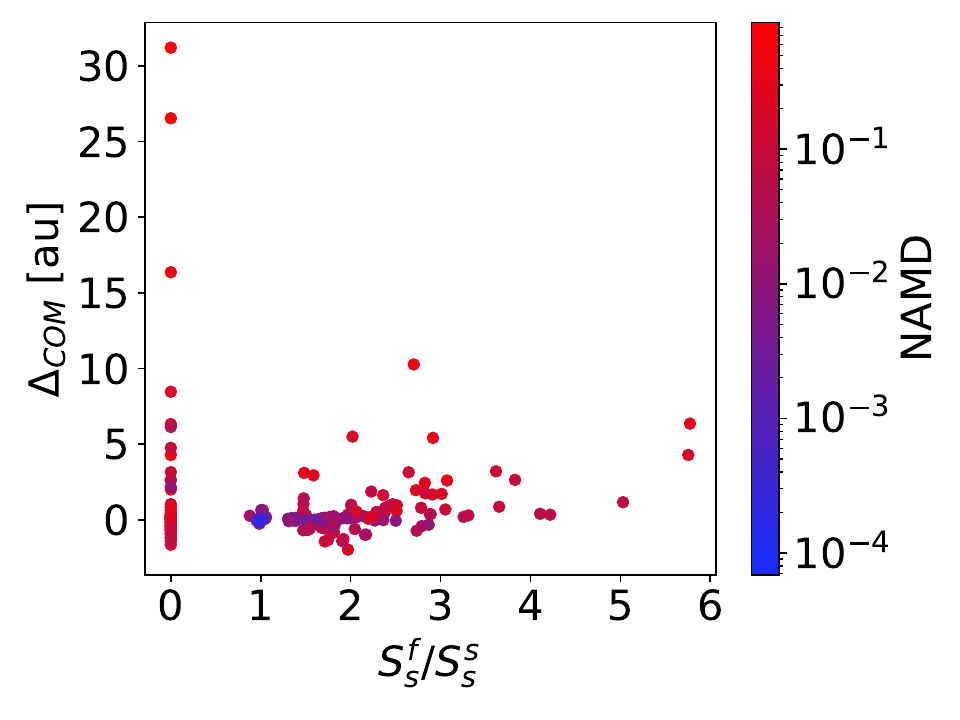}
    \;\;
    \includegraphics[width=0.95\columnwidth]{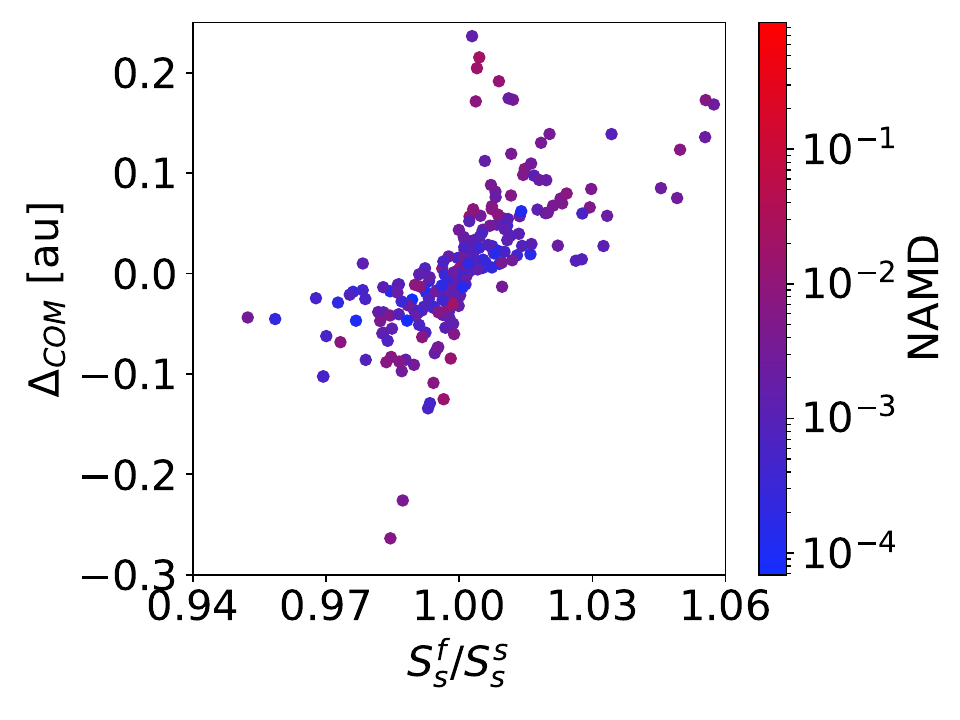}
    \vspace{-0.3cm}
    \caption{Population B: $S_m^f/S_m^s$ as a function of $S_s^f/S_s^s$ (top left panel) with a zoom-in of the region where both ratios are equal or very close to unity (top right panel) and $\Delta_{\rm CoM}$ as a function of $S_s^f/S_s^s$ (bottom left panel) with a zoom-in of the region where $S_s^f/S_s^s$ is equal or very close to unity and $\Delta_{\rm CoM}$ is equal or very close to zero (bottom right panel). Disrupted systems are not displayed in this figure as it was not possible to compute their NAMD and $\Delta_{\rm CoM}$.}
    \label{SsvsSmCOMDWD3}
\end{figure*}

\begin{figure}
\centering
    \includegraphics[width=\columnwidth]{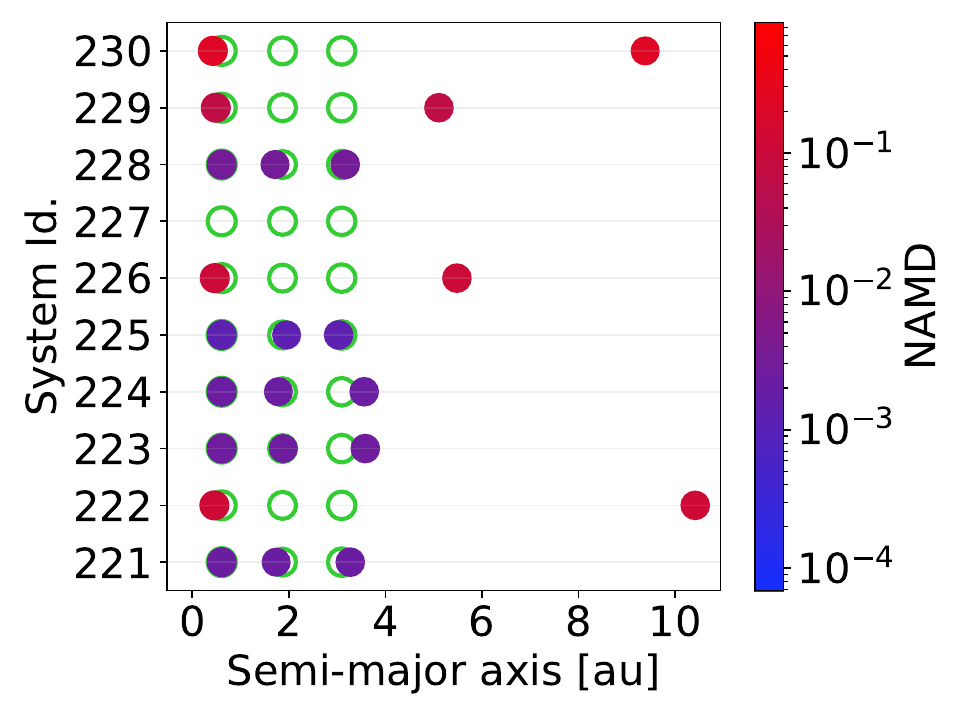}
    \includegraphics[width=\columnwidth]{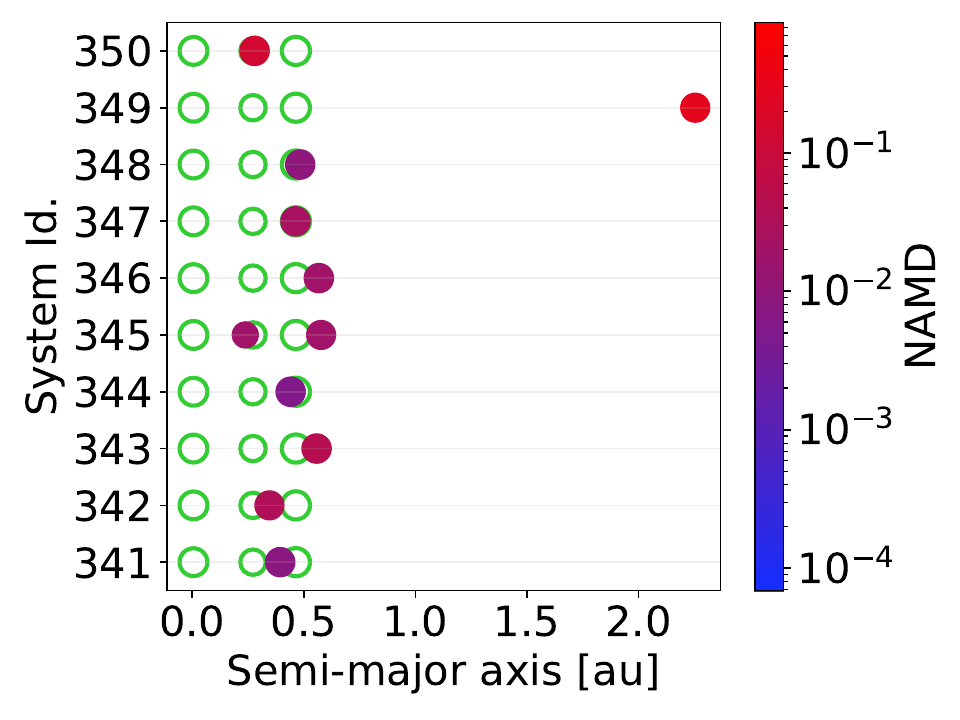}
    \vspace{-0.6cm}
    \caption{DWD$^*_3$ systems (Population B): initial and final $a$  (empty green circles, filled points, respectively). Systems IDs \#221-230 (on the vertical axis, top panel), System IDs \#341-350 (on the vertical axis, bottom panel). Each point represents a planet and its size is proportional to the planetary mass; the color-map gives a measure of the degree of stability. Note that in some systems the green circles overlap with the filled points.}
    \label{posDWD3-221-230&341-350}
\end{figure}

The systems in Population B that do not experience any catastrophic event (49\% of the simulated architectures) have $S_m^f/S_m^s=1$ and $S_s^f/S_s^s\simeq1$ (top right panel in Fig. \ref{SsvsSmCOMDWD3}). These systems remain dynamically cold, experience an ordered evolution, and their final architectures do not change significantly from the initial ones. The systems that are dynamically hot have both ratios greater than unity because the planets increase their mutual distance during the phase of instability and either the total mass of the planetary system decreases or the most massive planet accretes one of the smaller ones. Excluding the cases with only one surviving planet, which translates in $S_s^f/S_s^s=0$, all dynamical hot cases except one have $S_s^f/S_s^s>1$. The system in question is system ID \#345 that has a more compact final architecture (bottom panel in Fig. \ref{posDWD3-221-230&341-350}).

When we focus on the information given by the $\Delta_{\rm CoM}$, we immediately see that the stable systems, which have $S_{s,m}^f/S_{s,m}^s\simeq1$, do not experience a significant CoM shift (i.e., $|\Delta_{\rm CoM}|< 0.3$, bottom right panel in Fig. \ref{SsvsSmCOMDWD3}). 
The majority of systems with $S_s^f/S_s^s>1$, on the other hand, have $\Delta_{\rm CoM}>0$. As an illustrative example, systems ID \#222 and ID \#230 experience both one ejection, leading to the increase of the orbital separation among the surviving planets (top panel in Fig. \ref{posDWD3-221-230&341-350}). The CoM of these systems shifts outward since one of the surviving planets moves on a significantly wider orbit. Systems with only one surviving planet have $S_s^f/S_s^s=0$ and the shift in the CoM is inward if the surviving planet ends up orbiting closer to the binary, or outward if it moves to a wider orbit compared to its initial one. 
Figure \ref{posDWD3-221-230&341-350} also shows an example of a system (ID \#227) that is completely disrupted by the combination of instabilities and catastrophic events and for which the NAMD can not be computed.

\subsection{Population C}

\begin{figure}
   \centering
   \includegraphics[width=\hsize]{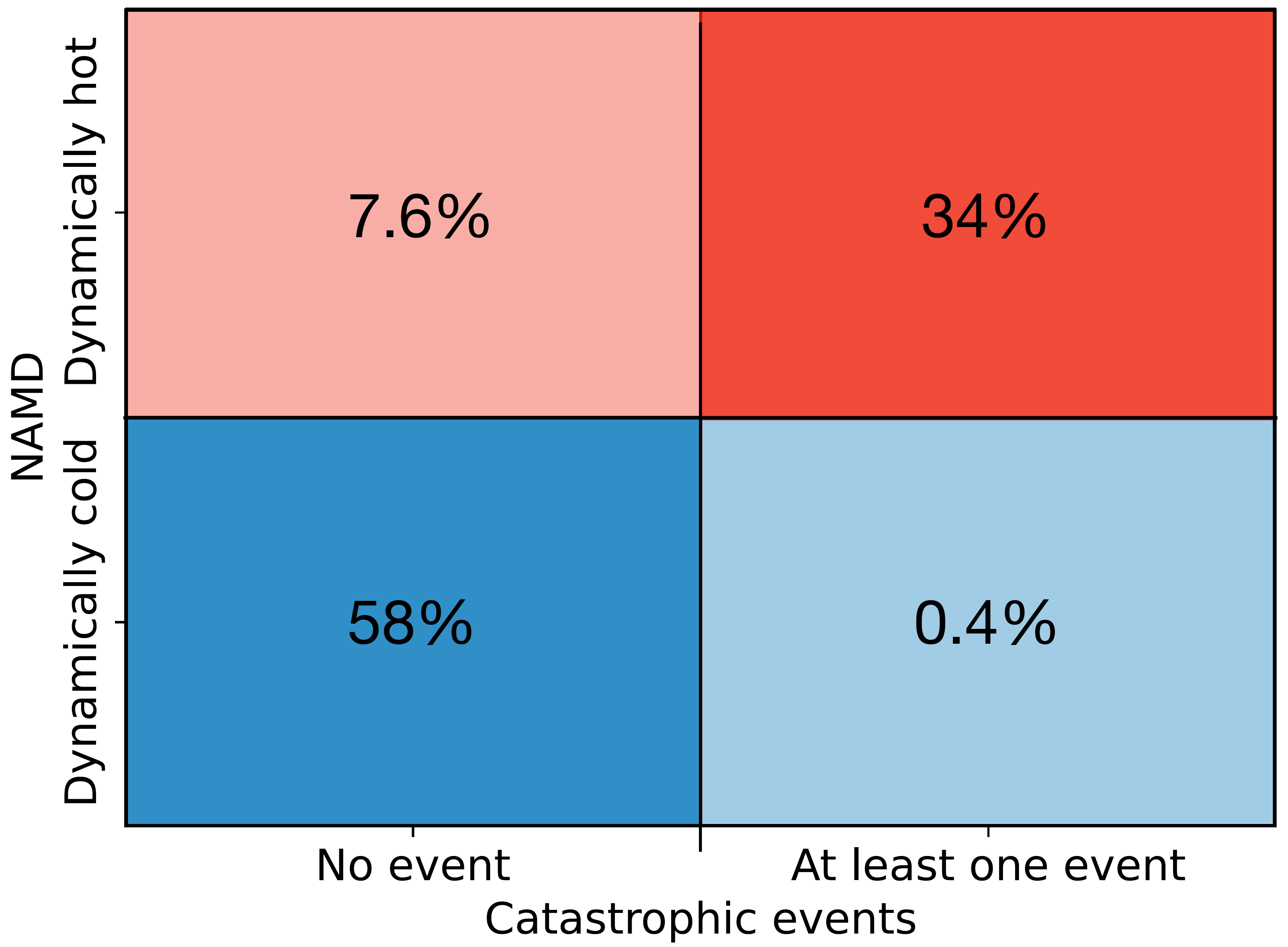}
      \caption{Population C: schematic representation of the percentage of simulated planetary systems (500 simulations) based on two specific properties: whether they experience or not at least one catastrophic event (ejection/scattering/collision) and whether they are dynamically cold or hot. The number in each box is related to the percentage of systems which have the combination of parameters specified on the x an y axes. The 34\% of systems displayed in upper right corner include the 12.2\% of systems that are disrupted (see text for details).}
         \label{heatmapC}
\end{figure}

\begin{figure*}
    \centering
    \includegraphics[width=0.95\columnwidth]{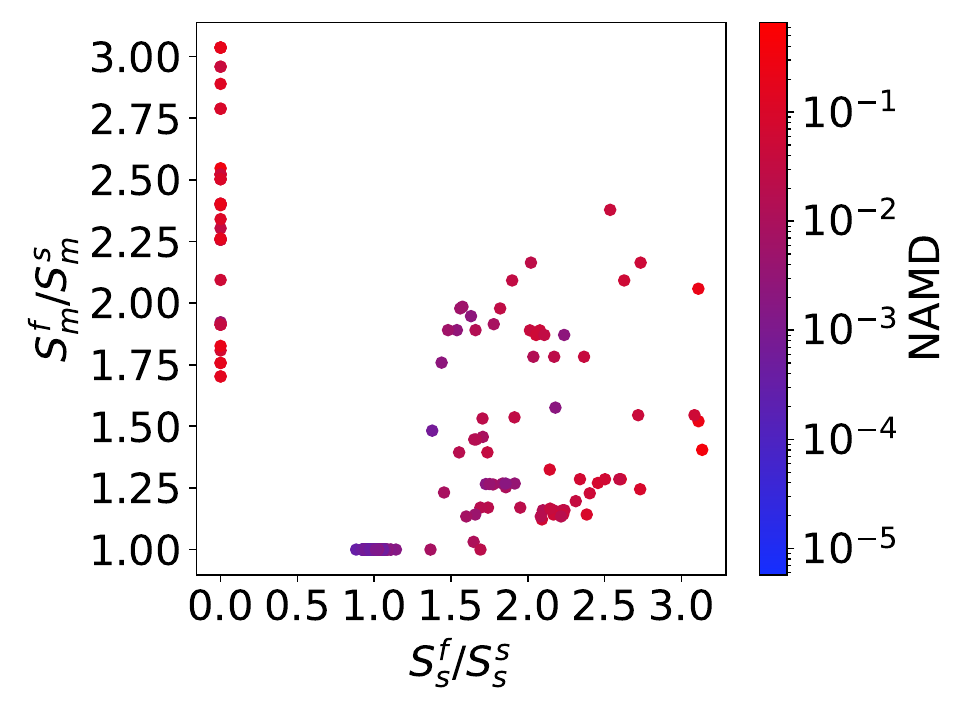}
    \;\;
    \includegraphics[width=0.95\columnwidth]{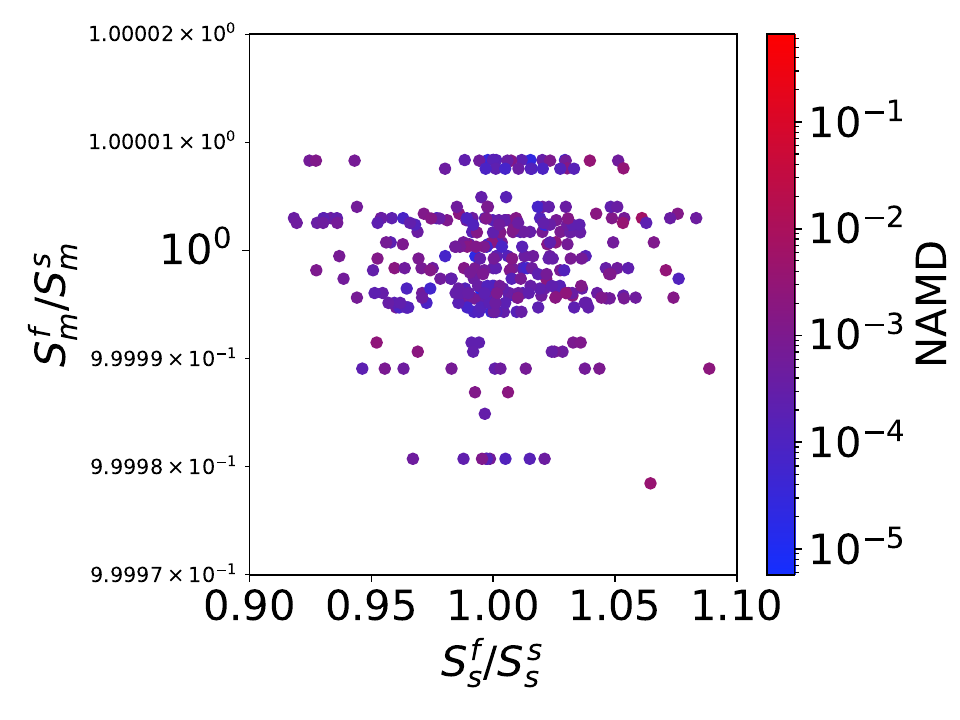}
    \;\;
    \includegraphics[width=0.95\columnwidth]{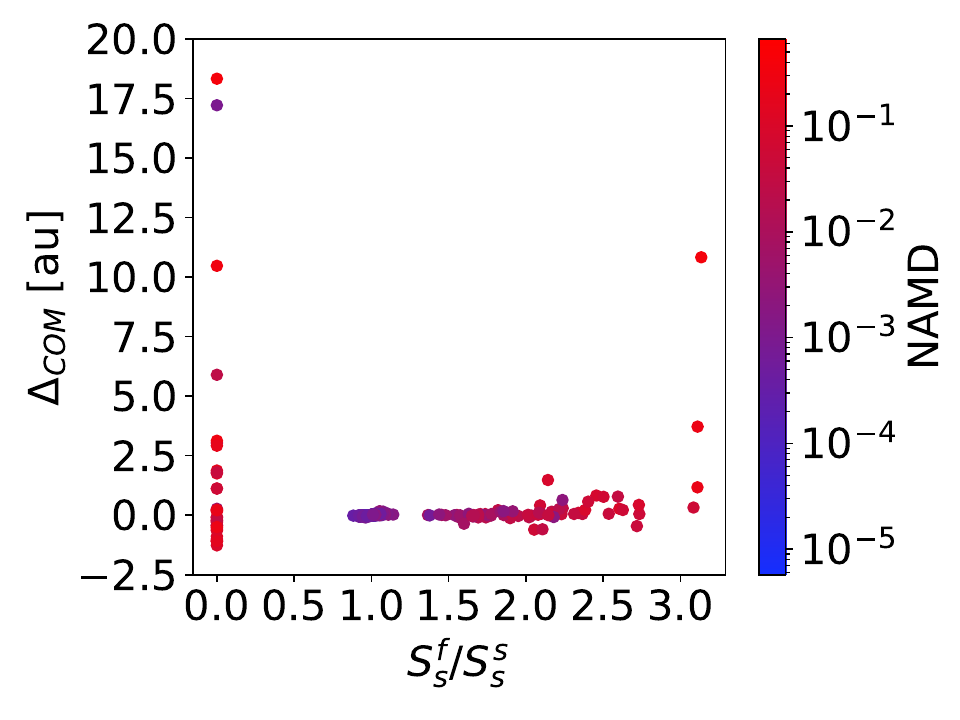}
    \;\;
    \includegraphics[width=0.95\columnwidth]{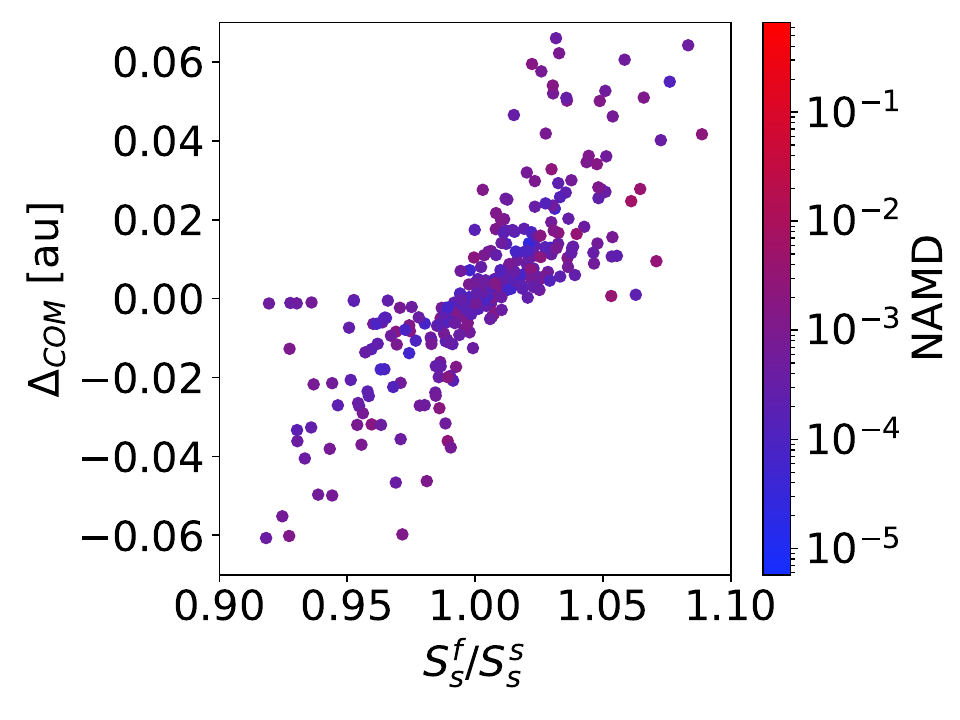}
    \vspace{-0.3cm}
    \caption{Population C: $S_m^f/S_m^s$ as a function of $S_s^f/S_s^s$ (top left panel) with a zoom-in of the region where both ratios are equal or very close to unity (top right panel) and $\Delta_{\rm CoM}$ as a function of $S_s^f/S_s^s$ (bottom left panel) with a zoom-in of the region where $S_s^f/S_s^s$ is equal or very close to unity and $\Delta_{\rm CoM}$ is equal or very close to zero (bottom right panel). Disrupted systems are not displayed in this figure as it was not possible to compute their NAMD and $\Delta_{\rm CoM}$.}
    \label{SsvsSmCOMDWD3_resonant}
\end{figure*}

Population C corresponds to the resonant counterpart of Population B. Compared to Population B, here we find that a larger fraction of systems remains dynamically cold over the simulated timescales (Fig. \ref{heatmapC}), with only 41.6\% of systems ending up dynamically hot, compared to the 72\% of Population B. However, compared to Population B, we have a slightly higher fraction of disrupted systems (12.2\% compared to 9\%), which in all cases except one experienced at least a close approach with the central binary.
In the sampled parameter space, the majority (65.6\%) of the systems does not experience catastrophic events, 27.2\% of systems experience ejections/scatterings: in 11.2\% of cases alone, while in 16\% of cases in combination with other catastrophic events (Tab. \ref{tabEventiPopAB}).
Normalizing the occurrence rates of catastrophic events to the frequency of dynamical hot systems, we find that 81.4\% end up loosing planets through catastrophic events (Tab. \ref{tabeventiNAMDmagAB}).
In terms of the metrics $S_m^f/S_m^s$, $S_s^f/S_s^s$ and $\Delta_{\rm CoM}$, we find an equivalent behavior as in Population B. The systems that become dynamically excited and experience planet loss, end up with an increased mass concentration in the most massive planet and with an increased orbital spacing among survived planets (left panels in Fig. \ref{SsvsSmCOMDWD3_resonant}). Dynamically cold systems, instead, have $S_m^f/S_m^s\simeq1$, $S_s^f/S_s^s\simeq1$ and $\Delta_{\rm CoM}\simeq0$ (right panels in Fig. \ref{SsvsSmCOMDWD3_resonant}), meaning they preserve their initial 2:1 resonance.

\subsection{Population D}

\begin{figure}
   \centering
   \includegraphics[width=\hsize]{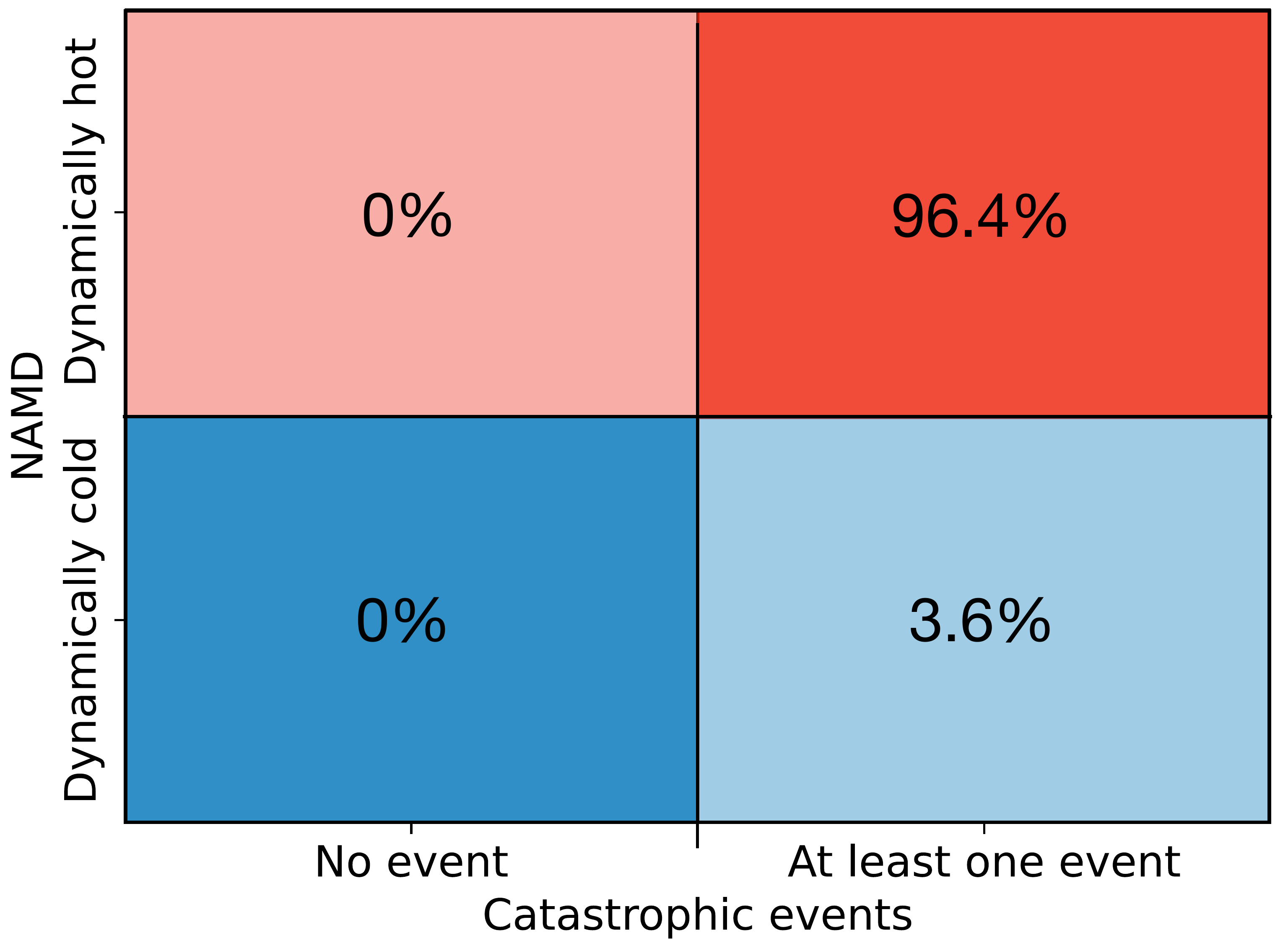}
      \caption{Population D: schematic representation of the percentage of simulated planetary systems (500 simulations) based on two specific properties: whether they experience or not at least one catastrophic event (ejection/scattering/collision) and whether they are dynamically cold or hot. The number in each box is related to the percentage of systems which have the combination of parameters specified on the x an y axes. The 96.4\% of systems displayed in upper right corner include the 17.2\% of systems that are disrupted (see text for details).}
         \label{heatmapD}
\end{figure}

With Population D we investigate the stability of five-planet systems in the NMS resonance (i.e., combination of 4:3 and 5:3 resonances). Compared to all the other populations, we find that all planetary systems undergo catastrophic events under the adopted priors and the majority of systems (96.4\%) ends up dynamically hot (Fig. \ref{heatmapD}). Nevertheless, 82.8\% of planetary systems have at least one surviving planet. The most common outcome for Population D planetary systems is to undergo combinations of different catastrophic events (Tab. \ref{tabeventimolteplicita}) and lose, in the majority of cases (58\%) two of their initial planets (Tab. \ref{tab.ejectedbodAB}). 

\begin{figure}
    \centering
    \includegraphics[width=\columnwidth]{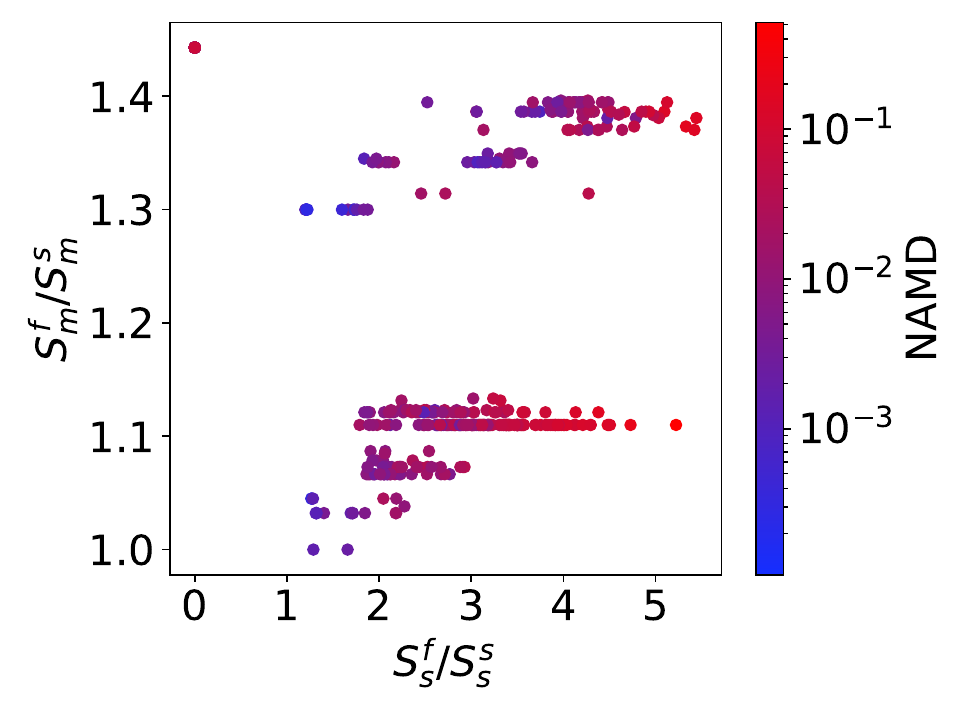}
    \vspace{-0.6cm}
    \caption{Population D: $S_m^f/S_m^s$ as a function of $S_s^f/S_s^s$. Disrupted systems are not displayed in this figure as it was not possible to compute their NAMD and $\Delta_{\rm CoM}$.}
    \label{SsvsSmDWD4_resonant}
\end{figure}

\begin{figure*}
    \centering
    \includegraphics[width=0.95\columnwidth]{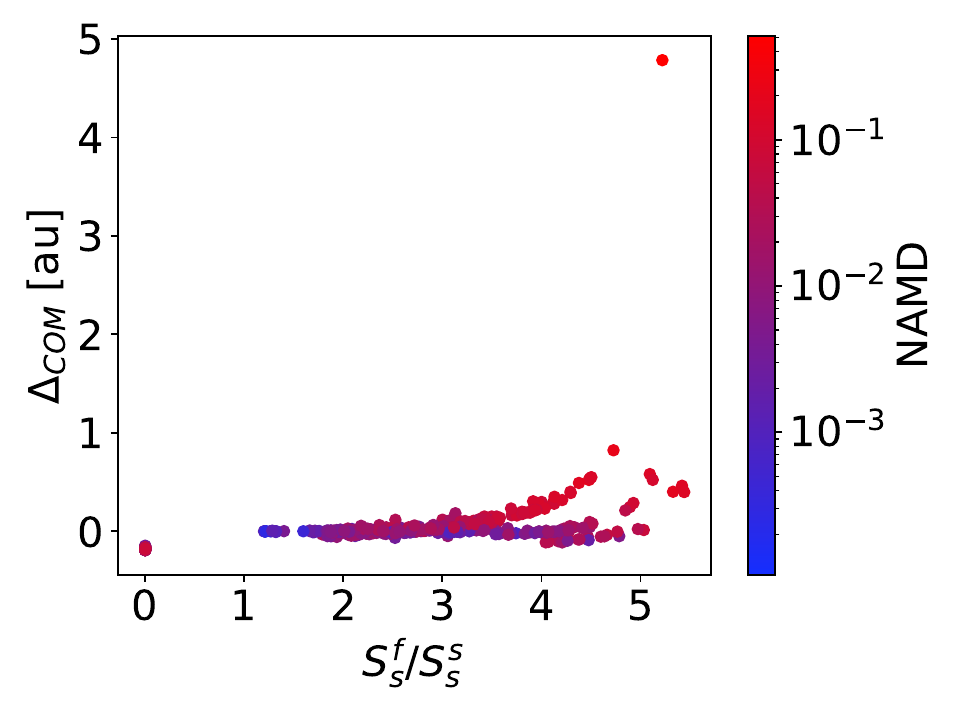}
    \;\;
    \includegraphics[width=0.95\columnwidth]{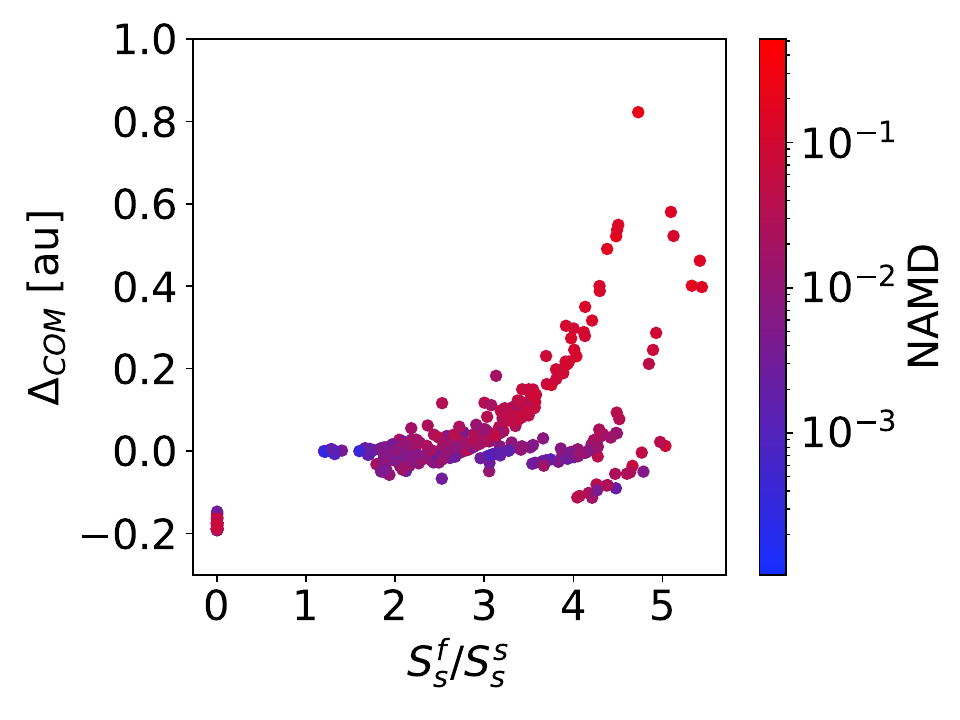}
    \vspace{-0.3cm}
    \caption{Population D: $\Delta_{\rm CoM}$ as a function of $S_s^f/S_s^s$ (left panel) with a zoom-in of the region where $\Delta_{\rm CoM}<1$ (right panel). Disrupted systems are not displayed in this figure as it was not possible to compute their NAMD and $\Delta_{\rm CoM}$.}
    \label{SsvsCOMDWD4_resonant}
\end{figure*}

\begin{figure}
\centering
    \includegraphics[width=\columnwidth]{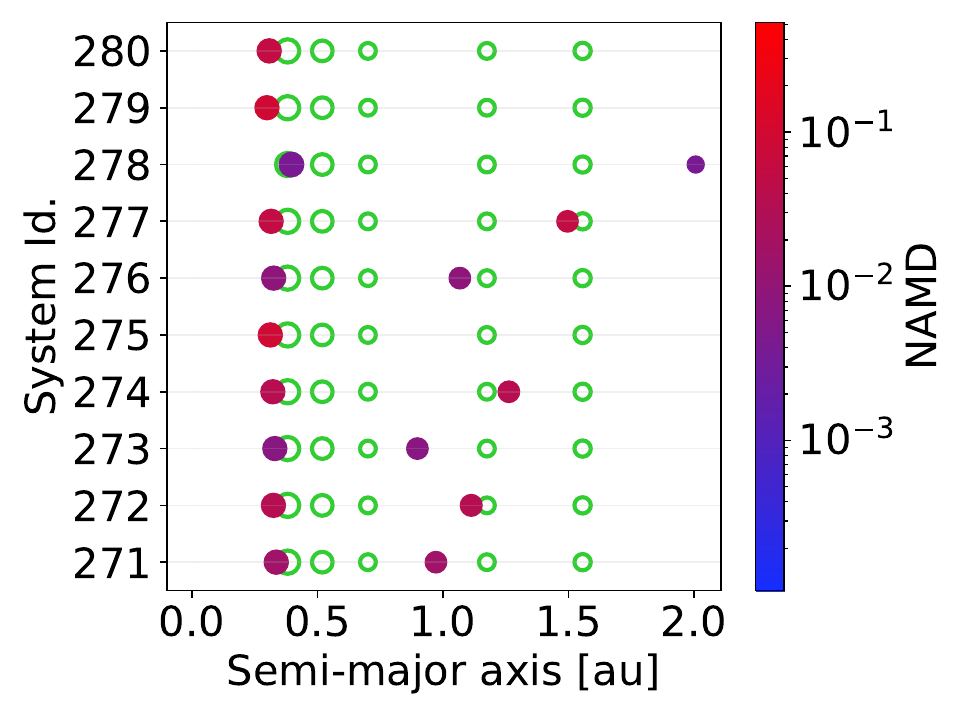}
    \vspace{-0.6cm}
    \caption{DWD$^*_4$ systems (Population D): initial and final $a$  (empty green circles, filled points, respectively). Systems IDs 271-280 (on the vertical axis). Each point represents a planet and its size is proportional to the planetary mass; the color-map gives a measure of the degree of stability. Note that in some systems the green circles overlap with the filled points.}
    \label{posDWD4-271-280}
\end{figure}

All planetary systems with more than one surviving planet have $S_m^f/S_m^s>1$, $S_s^f/S_s^s>1$ (Fig. \ref{SsvsSmDWD4_resonant}) and the center of mass experiences both positive or negative shifts. All the planetary systems that end up with only one surviving planet have $\Delta_{\rm CoM}<1$ (Fig. \ref{SsvsCOMDWD4_resonant}, the points with $S_s^f/S_s^s=0$), meaning that the surviving planet does not get scattered outward, as it happened instead with the other three populations. Depending on which one of the initial planets survived and whether it accreted other planets, its final mass is always between 1 \Mj\ and 1.053 \Mj, hence the $S_m^f/S_m^s$ metric gives similar values and the points are overlapping in Figure \ref{SsvsSmDWD4_resonant}. Illustrative examples of this scenarios are shown by system IDs \#275, 279, 280 in Figure \ref{posDWD4-271-280}.


\subsection{Dependence of the dynamical evolution on the initial multiplicity}\label{subsec:dependanceonmulti}

When we divide the planetary systems based on their initial number of planets, the NAMD value of the planetary systems shows a dependence on the multiplicity (Tab. \ref{tabNAMDpianeti}, where we report Populations A, B, C only, for which we sampled the initial number of planets). Systems initially hosting two planetary companions are more likely to remain in a stable configuration and preserve their original architectures over the simulated timescales. These systems preserve their multiplicity even when going through an excited phase as showcased by Population B in Tab. \ref{tabNAMDpianeti}).

A large fraction of planetary systems born with more than two planets (e.g., system IDs \#222, 226, 229, 230 in Fig. \ref{posDWD3-221-230&341-350} and system IDs \#442, 449, 450 in Fig. \ref{posPopA_merged}) can still experience an ordered evolution, yet the more planets initially populate a system, the more probable it is for it to go through an unstable phase which could also disrupt it in the most catastrophic scenarios (Tab. \ref{tabNAMDpianeti}). Under the adopted priors, the fraction of systems undergoing catastrophic events grows from 12.9\% to 45.2\% when going from three-planet to four-planet systems for Population A, from 54.2\% to 79.1\% for Population B, and from 51.6\% to 95\% for Population C (Tab. \ref{tabeventimolteplicita}). Likewise, in the case of Population A the frequency of disrupted systems grows from 0\% to 0.9\%, while in the case of Population B and C this frequency grows from 8.1\% to 18.2\% and from 17.7\% to 40\%, respectively (Tab. \ref{tabNAMDpianeti}).

Consistently with the explanation proposed by \cite{turrini2020} for the NAMD-multiplicity anti-correlation in multi-planet systems around Main Sequence stars, our results clearly show that, when the multiplicity of the planetary system increases so does the likelihood that the system enters an excited phase and, as a consequence, experiences ejections/scatterings and/or collisions. Reinforcing this trend, we find that all the five-planet systems, that belong to Population D, experience planet loss.

\begin{table*}[b]
    \caption{Percentage of Population A, B and C planetary systems with the NAMD value in a given range.}
    \centering
    \begin{tabular}{l|c|c|c||c|c|c||c|c|c}
        \hline
        \hline
        & \multicolumn{3}{c||}{Pop. A} & \multicolumn{3}{c||}{Pop. B} & \multicolumn{3}{c}{Pop. C}\\
        \hline
        & $N_s=2$ & $N_s=3$ & $N_s=4$ & $N_s=2$ & $N_s=3$ & $N_s=4$ & $N_s=2$ & $N_s=3$ & $N_s=4$\\
        \hline
         NAMD<$1.3\times10^{-3}$ & 100\% & 87.8\% & 61.3\% & 73.7\% & 21.6\% & 10.9\% & 93\% & 39.6\% & 5\% \\
         NAMD>$1.3\times10^{-3}$ & 0\% & 12.2\% & 37.8\% & 26.3\% & 70.3\% & 70.9\% & 7\% & 42.7\% & 52.5\% \\
         Disrupted systems & 0\% & 0\% & 0.9\% & 0\% & 8.1\% & 18.2\% & 0\% & 17.7\% & 42.5\% \\
         \hline    
    \end{tabular}
    \label{tabNAMDpianeti}
    \tablefoot{The percentages are categorized based on the initial number of planets (see Tab. \ref{tabpianetiiniziali} to know how many simulations had initially 2, 3 and 4 planets).}
\end{table*}

\begin{table}[h]
    \caption{\label{tabeventimolteplicita}Percentage of the Population A, B and C planetary systems which experienced catastrophic events.}
    \centering
    \begin{tabular}{l|c|c|c}
        \hline
        \hline
        \multicolumn{4}{c}{Pop. A}\\
        \hline
         & $N_s=2$ & $N_s=3$ & $N_s=4$\\
         \hline
         No catastrophic events & 100\% & 87.1\% & 54.8\%\\
         Ejections/scatterings only & 0\% & 6.7\% & 21.3\%\\
         Collisions$_\text{p-p}$ only & 0\% & 6.2\% & 11.7\%\\
         Close approach to the binary & 0\% & 0\% & 3.5\%\\
         Multiple catastrophic events & 0\% & 0\% & 8.7\%\\
         \hline
         \hline
         \multicolumn{4}{c}{Pop. B}\\
         \hline
         & $N_s=2$ & $N_s=3$ & $N_s=4$\\
         \hline
         No catastrophic events & 100\% & 45.8\% & 20.9\%\\
         Ejections/scatterings only & 0\% & 23.2\% & 28.2\%\\
         Collisions$_\text{p-p}$ only & 0\% & 18.4\% & 17.3\%\\
         Close approach to the binary & 0\% & 0.3\% & 0\%\\
         Multiple catastrophic events & 0\% & 12.3\% & 33.6\%\\
         \hline
         \hline
         \multicolumn{4}{c}{Pop. C}\\
         \hline
         & $N_s=2$ & $N_s=3$ & $N_s=4$\\
         \hline
         No catastrophic events & 100\% & 48.4\% & 5\%\\
         Ejections/scatterings only & 0\% & 19.2\% & 15\%\\
         Collisions$_\text{p-p}$ only & 0\% & 11.2\% & 15\%\\
         Close approach to the binary & 0\% & 0\% & 0\%\\
         Multiple catastrophic events & 0\% & 21.2\% & 65\%\\
         \hline             
    \end{tabular}
    \tablefoot{The Table is divided by taking into account the initial number of planets (see Tab. \ref{tabpianetiiniziali} to know how many simulations had initially 2, 3 and 4 planets).}
\end{table}

\section{Discussion} 
\label{discussion} 

\subsection{Observational implications of this stability study}

Our results provide insight on the survival likelihood of second-generation multi-planet systems and, consequently, on the prospective GW detection of systems possessing architectures similar to those explored in this study. Specifically, over the sampled parameter space and simulated timescales, the joint analysis of all the populations shows that 92.3\% of the 2500 simulated planetary systems (i.e., 500 systems for each DWD binary) are not disrupted during their evolution. Furthermore, 39.4\% of these 2500 systems preserves their initial multiplicity, while 52.9\% lose at least one planet without being disrupted. Excited systems, i.e. those with NAMD>$1.3\times10^{-3}$, are more likely to have undergone unstable phases and to have experienced catastrophic events, which in the majority of cases are ejections (Tab. \ref{tabEventiPopAB}; Fig. \ref{heatmapA}, \ref{heatmapDWD3}, \ref{heatmapC} and \ref{heatmapD} for Population A, B, C and D, respectively).
When we combine the four Populations and analyze them as a single one, we see that in 15.9\% of cases the planetary systems experienced scatterings/ejections, while in 17.8\% of cases the systems underwent multiple catastrophic events. In these extreme scenarios, where multiple catastrophic events take place, the planetary systems are generally disrupted. Disruptions, however, affect 0.2\% of the systems in Population A, 9\% of the systems in Population B, 12.2\% of systems in Population C and 17.2\% of systems in Population D, corresponding to 7.7\% of the total 2500 simulated systems. The reason why disruptions are more frequent in Population B and C rather than Population A is that Population B and C have more massive planets orbiting around a more massive binary, resulting in stronger gravitational perturbations between the bodies in the systems. The higher disruption fraction in Population D instead reflects the initial higher multiplicity, which is five for all these simulated systems, and the effects of the perturbations are enhanced by the more packed architecture. 

Concerning the difference between an initial non-resonant architectures and resonant ones, we can compare Population B with Population C, which employ the same binary system and sample the same range of planet masses, with the difference that in Population C the planets are initially in a 2:1 resonance. Under the adopted priors, more orbital architectures remain dynamically cold when they are in resonance (Fig. \ref{heatmapDWD3} and \ref{heatmapC}). When dividing the systems based on their initial multiplicity we also see that the fraction of stable two and three-planet systems is higher when such systems are in resonance (Tab. \ref{tabeventiNAMDmagAB}). However, in both populations B and C, the majority of three and four-planet systems experience catastrophic events and actually the fraction of four-planet systems that are disrupted is larger when these systems are in resonance. Moreover, in Population D, where we employ five-planet systems that are initially in the NMS resonance, we find that the majority (96.4\%) of systems ends up dynamically hot and all systems lose at least one planet, breaking the resonant chain (Tab. \ref{tabEventiPopAB}, \ref{tab.ejectedbodAB}).

Looking at both the resonant and the non-resonant scenarios, NAMD and the catastrophic events that a system experiences are linked to its initial multiplicity. Systems with higher multiplicity are more likely to undergo unstable phases after being excited (Tab. \ref{tabNAMDpianeti} and \ref{tabeventimolteplicita}). Systems with initial planet multiplicity of two never experience catastrophic events, even the 26.3\% of Population B and the 7\% of Population C two-planet systems that end up in excited architectures. Among all simulated two-planet systems, this scenario happened 7\% of times. Furthermore, 30.5\% of planetary systems with initially three, four or five planets end up with two surviving ones, increasing our initial two-planet population by 122\% and creating a degeneracy in knowing the past of potentially observed two-planet systems.

\begin{figure}
\centering
    \includegraphics[width=\columnwidth]{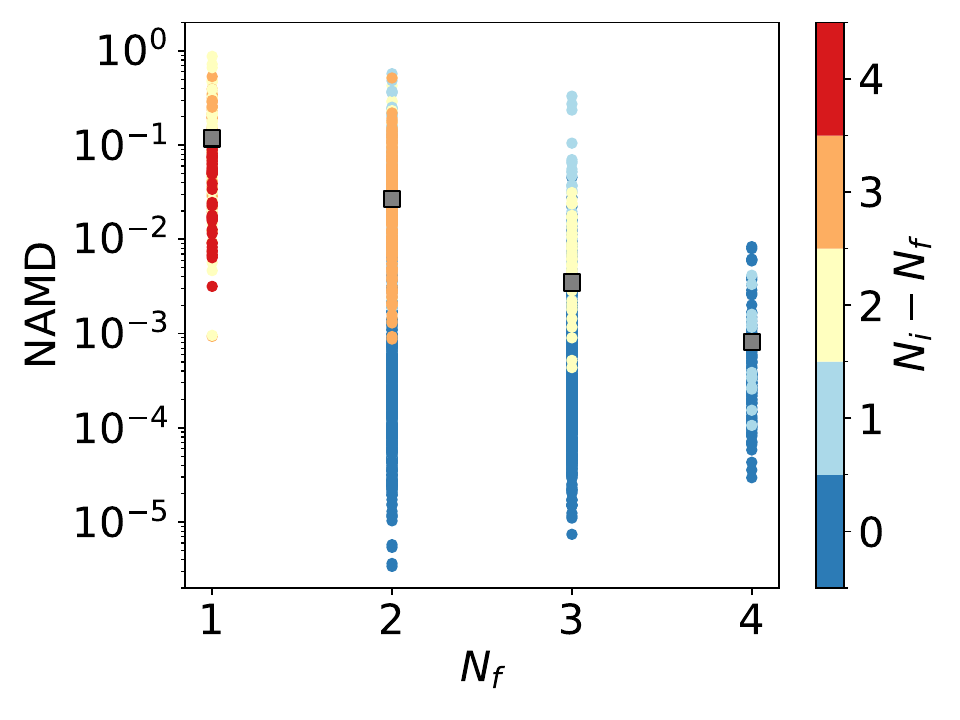}
    \vspace{-0.6cm}
    \caption{Final multiplicity versus NAMD for all the planetary systems considered in this study (colored circles) and for the average values of each multiplicity bin (gray squares). The colorbar indicates the number of lost planets for each system.}
    \label{NAMD-multiplicity}
\end{figure}

To this regard, in Figure \ref{NAMD-multiplicity} we show the NAMD as a function of the final number of planets, $N_f$, represented as circles and color-coded by the number of lost planets. The gray squares indicate the mean NAMD value for each $N_f$ bin. We find an anti-correlation between the NAMD and the final system multiplicity, in agreement with the results of \cite{turrini2020}.

Furthermore, as proposed also by \cite{turrini2020} for Main-Sequence stars, our results suggest that we can assess the likelihood that such observed two-planet systems had higher primordial multiplicity by quantifying their NAMD values after we characterize the orbits of their planets. If the NAMD $> 1.3\times10^{-3}$, i.e., the system is more dynamically excited than the Solar System, it is plausible that the system itself was originally more populated since in the majority of cases the two-planets systems that do not experience catastrophic events remain dynamically colder than this threshold value. This insight opens up to possibility to infer the loss of planets within second-generation systems hosting giant planets. As the results of \cite{Ledda2023} highlight that giant planets can form only over limited temporal intervals in second-generation discs, constraining how frequent these giant planets are can provide independent constraints on the timescale and efficiency of the planet formation process. It must be pointed out, however, that this observational outlook does not account for the possible presence within the system of first-generation planets which survived the evolution of the inner binary \citep{Columba2023}. Disentangling this possible source of degeneracy will require dedicated studies of planet formation within second-generation discs in the presence of a wide orbit planet \citep{Columba2023}, and of the orbital stability of the newly formed systems, that are beyond the scope of this paper.

\subsection{The LISA framework}

In the context of the future ESA-LISA space mission, we now focus on DWD$^*_3$ and DWD$^*_4$, as they belong to the LISA DWD synthetic Population. This means that, when such systems are placed within the context of the Milky Way’s history, the frequency $f_{\rm GW}$ of their gravitational wave falls within the LISA band \citep{Korol2017,Korol2019,AmaroSeoane2022:WP}.
For such, any sub-stellar object (SSO) orbiting these binaries might have the potential to be detected, provided that the signal-to-noise of the binary $S_{\rm DWD}$ is larger than $\sim$10 \citep[][their figure A.1]{dani2019} and that the orbital period and mass of the companion are optimal for causing a detectable modulation of the gravitational wave signal.  

We note that the considerations we draw in the following, concerning the possible GW detection of our simulated systems, are purely qualitative and model dependent, since detailed assessment of their detectability is  beyond the scope of this article. Dedicated studies would need to be performed to test whether LISA can hear the systems here studied. Multi-planet detection with LISA is a topic that has yet to be investigated: the presence of more than one planet would change the gravitational perturbation on the binary, and consequently also the final waveform compared to what presented in \cite{dani2019} and \cite{Katz2022}, based on a single-planet approach.

\begin{figure}
    \centering
    \includegraphics[width=\hsize]{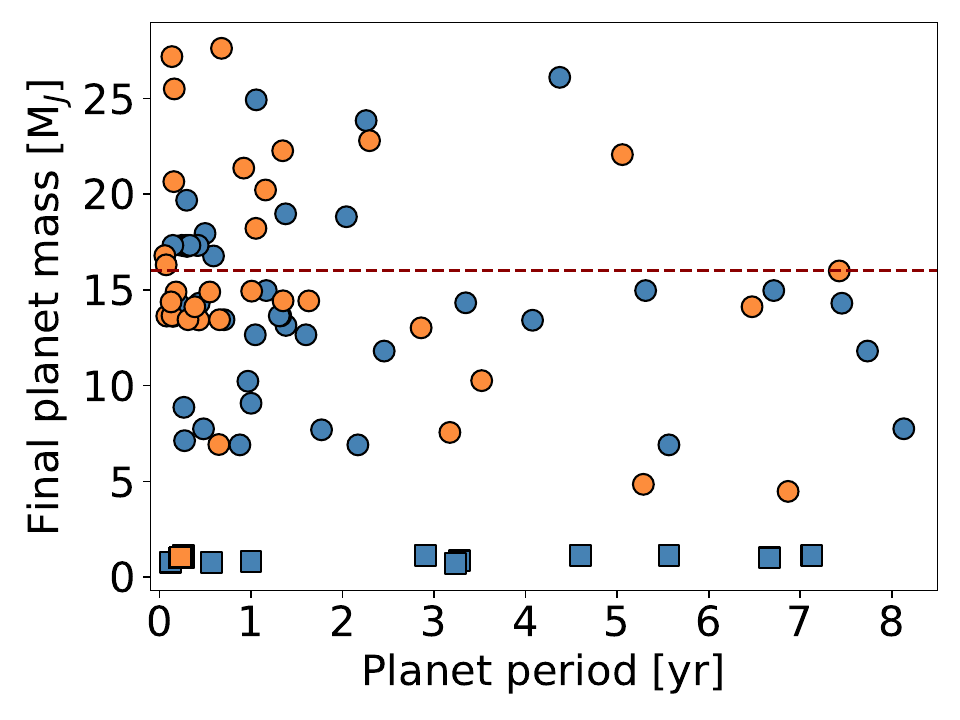}
    \vspace{-0.6cm}
     \caption{Population B (blue circles), Population A (blue squares), Population C (orange circles) and Population D (orange squares) systems: Final planet mass vs final planet period. The red dashed line at \Mp\ = 16 \Mj\ represents the threshold that separates planets from brown dwarfs (BDs) as discussed in \cite{Spiegel2011}. We end up with BDs because some planets undergo collisional merge and the resulting total planet mass is greater than 16 \Mj. We note that some points are overlapping with each other, especially the ones relative to Population D.}
    \label{fig:MpVSs_DWD3}
\end{figure}

For what it concerns those systems that end up with only one SSO, most of the bodies (i.e., a total of 121 planets and 28 brown dwarfs, Fig. \ref{fig:MpVSs_DWD3}) have masses \Mp\ greater than 4 \Mj\ and a period less than half the observing time of the LISA nominal mission (4 years, assuming 100\% duty cycle). Following the results by \cite{Katz2022} these single planet systems have the potential to be detected, pending their distance, sky-location, polarization angle \citep{Robson2018}. 
Overall, the GW signal-to-noise ratio of a DWD is determined by these parameters, together with the binary’s chirp mass, orbital period, and eccentricity, if any. Since the detection of $Magrathea$ planets relies on the perturbation of the binary’s center of mass, their detectability primarily depends on the planetary mass and orbital separation. We refer the reader to \cite{dani2019}, their Appendix A, for a detailed study of the parameter space covered by $Magrathea$ planets and brown dwarfs.

Concerning those systems with more than one planet, we can assess as a first approximation the detectability of each planetary system around DWD$_3^{*}$ and  DWD$_4^{*}$, by measuring the gravitational wave frequency shift amplitude produced by the planets in their systems.  
Following the syntax of \cite{dani2019}, the motion of the DWD around the center of mass of a three body system modulates the GW frequency through the Doppler effect where the resulting frequency observed by LISA is
\begin{equation}
 f_{\rm obs}(t) = \left( 1 + \frac{v_{\parallel}(t)}{c} \right) f_{\rm GW}(t) \,.
 \label{eq:Doppler}
\end{equation}

\noindent $v_\parallel = - K \cos\varphi(t) \,$ is the line of sight velocity of the DWD with respect to the common center of mass,  $f_{\rm GW} = 2 \cdot f_{\rm orb}$ is the GW frequency in the reference frame at rest with respect to the DWD center of mass and $f_{\rm orb}$ is the orbital frequency of the binary \citep{maggiore2008gravitational,postnov2014}. The parameter $c$ is the speed of light,  $\varphi(t)$ is the orbital phase (see \citealt{dani2019}, their Eq. 5),   and $K$ is the amplitude of the radial velocity signal caused by the presence of the third object orbiting the binary and it is defined as 

\begin{equation}
 K = \left(\frac{2\pi G}{P}\right)^{\frac{1}{3}} \frac{M}{(M_{\text{bin}}+M)^{\frac{2}{3}}} \sin i  \,,
 \label{eq:K}
 \end{equation}

\noindent where $P$, $M$ and $i$ are the third body period, mass and inclination, respectively.  $M_{\text{bin}}$ is the DWD total mass, and $G$ the gravitational constant. 
Given that the average $\cos\varphi(t) \sim 1$
the gravitational wave frequency shift amplitude for a system with one planet can be approximated from Eq \ref{eq:Doppler} as

\begin{equation}
    \Delta f = f_{\rm obs} - f_{\rm GW} \eqsim f_{\rm GW} \frac{K}{c}.
 \label{eq:multiplanet}
\end{equation}

\noindent Yet, in the event of a multi-planet system with $N$ planets, assuming that their orbital angular momentum vectors are isotropically distributed, i.e. the average of $\sin{i}= \nicefrac{\pi}{4}$, we can then approximate the global shift as 

\begin{equation}\label{deltaF}
 \Delta f =   \frac{\pi}{4}  \left(\frac{2\pi G}{c^3}\right)^{\frac{1}{3}} f_{\rm GW} \sum_{j=1}^N \left( \frac{M_j \cdot P_j^{-\frac{1}{3}}}{(M_{\text{bin}}+M_j)^{\frac{2}{3}}} \right)
\end{equation}

\begin{table}[t]
    \caption{\label{tempiDWD_FormationEvolution} Minimum and maximum times needed for DWD$^*_3$ and DWD$^*_4$ systems to form giant planets and to reach a stable orbital configuration, and time needed for the stars in the binary to merge.}
    \centering
    \begin{tabular}{l|c|c||c}
        \hline
        \hline
         & $t_{\text{min}}$ [Myr] & $t_{\text{max}}$ [Myr] & $t_{\text{merge}}$ [Myr]\\
         \hline
         DWD$^*_3$ & 0.65 & 10.9 & 150 \\
         DWD$^*_4$ & 1.00 & 10.6 & 430 \\
         \hline
         \hline
         Pop. B ID \#126 & 0.67 & 3.02 & 430 \\
         Pop. B ID \#235 & 2.65 & 5.00 & 430 \\         
         \hline                  
    \end{tabular}
    \tablefoot{These timescales take into account the time needed to form the first giant planet in the simulations of \cite{Ledda2023} (see its Tab. A.6 and A.9) and the time needed to reach a stable configuration in our simulations; $t_{\text{merge}}$ is taken from Tab. 1 of \cite{Ledda2023}. The minimum and maximum times (top) are calculated accounting for the whole set of performed simulations (500 for each DWD); The minimum and maximum times (bottom) are only relative to the specific systems ID \#126 and \#235, which experienced the largest positive CoM shift among the nominal, not resonant, simulations.}
    \label{tab:merge}
\end{table}

\begin{figure*}[t]
\centering
    \includegraphics[width=0.9\columnwidth]{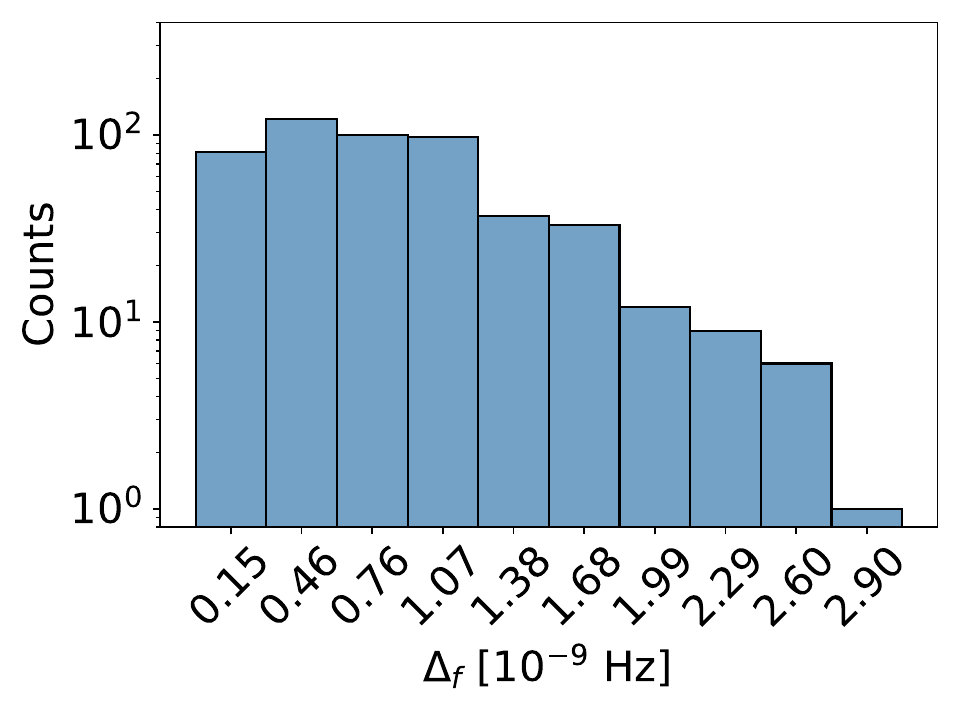}
    \includegraphics[width=0.9\columnwidth]{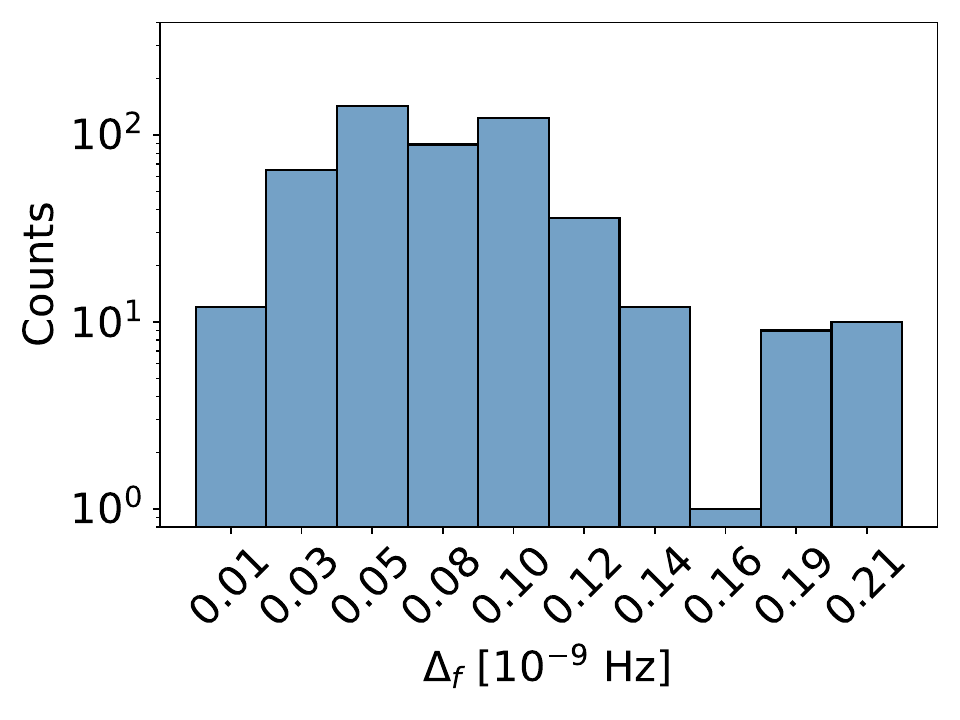}
    \includegraphics[width=0.9\columnwidth]{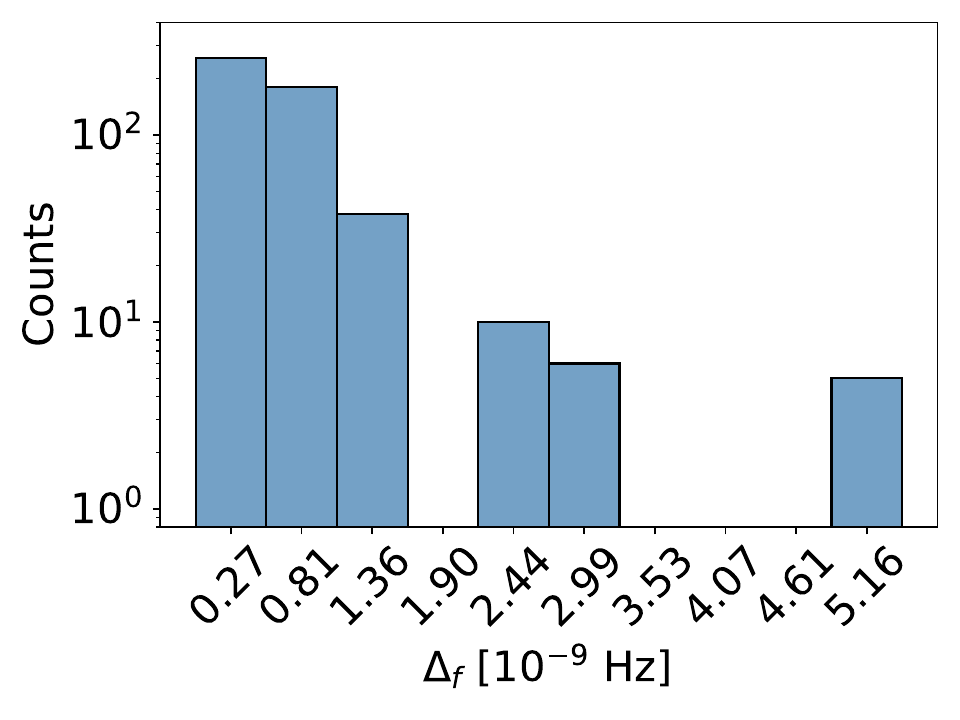}
    \includegraphics[width=0.9\columnwidth]{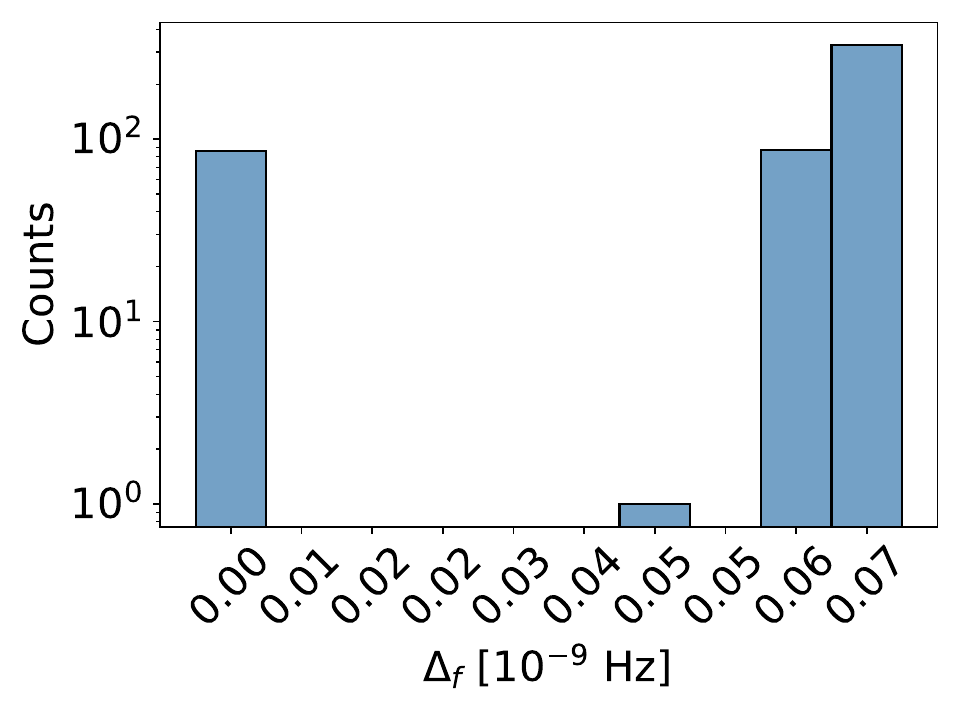}
    \vspace{-0.3cm}
     \caption{Population B (top left), Population A 
     (top right), Population C (bottom left) and Population D (bottom right) systems: global frequency shift $\Delta f$ computed with eq. \eqref{deltaF}.}
    \label{DeltaF_DWD}
\end{figure*}

\begin{figure}
    \centering
    \includegraphics[width=\hsize]{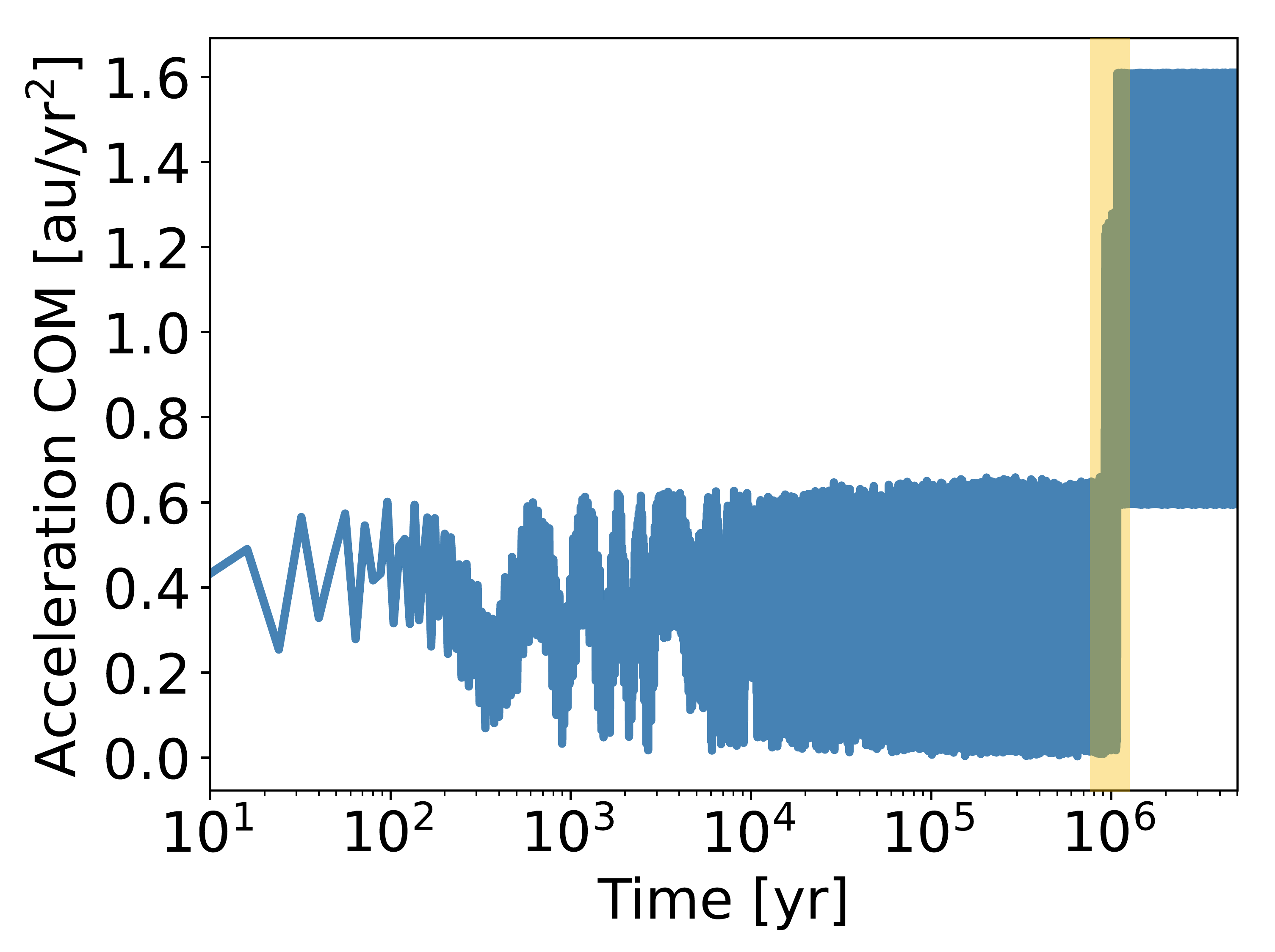}
    \vspace{-0.6cm}
     \caption{Acceleration of the planetary center of mas versus time for System ID \#235, which experienced one of the largest $|\Delta_{\rm CoM}|$. The yellow shaded region indicates when ejections took place in the simulation.} 
    \label{fig:COM_acceleration}
\end{figure}

For all the analyzed systems such shift is smaller than LISA frequency resolution of $\sim 7.92\times10^{-9}$ Hz for a 4 years mission. The majority of systems is also below the frequency resolution of 3.96$\times10^{-9}$ Hz (8 years mission), except for a handful of systems that belong to Population C (bottom left panel in Fig. \ref{DeltaF_DWD}). 

The gravitational wave frequency emitted by the DWD$_3^{*}$ and DWD$_4^{*}$ binaries ($f_{\rm GW}$ = 0.36 mHz, 0.32 mHz respectively) lie within the LISA band (i.e., 0.1 mHz–1 Hz, \cite{Amaro-Seoane2023}), meaning that their detection is possible, provided that the distance to the binaries is favorable.
However, the gravitational shift induced by the planets in each system would be too small to be detected by LISA in the majority of the cases.
The conclusions we can draw based on these results are two: (i) a specific waveform study, such as the one by \cite{Katz2022}, done for a single-planet system case, is needed to definitively test whether each system is realistically detectable; (ii) the shift was computed using the binary parameters right after the second common envelope phase. Yet, the binary will shrink over time, shortening the orbital period and hence increasing the orbital frequency. An increase in the orbital frequency would allow for a higher probability of detection of a third body perturbing the DWD \citep{dani2019}. 
For instance, in the case of a 2 planets system (like ID \#219 from Population B, characterized by $\Delta f = 1.50\times10^{-9}$ Hz), the binary period should shrink by a factor of 6 or 3 to make the frequency shift amplitude caused by the planets detectable for a 8 or 4 year mission, respectively. 
We note that we studied the stability of the systems formed around DWD$_3^{*}$ and DWD$_4^{*}$ over 10 Myr after the DWD formation of the double generate binary itself. 
In such a timescale, the period of the binary itself would have shorten by only $\sim$ 2.3 minutes and $\sim$ 1 min for DWD$_3^{*}$ and DWD$_4^{*}$, respectively\footnote{This example is meant only to provide an order of magnitude of the binary frequency change. It is computed based on Eq 9 of \cite{dani2019} and it is meant for a DWD alone, excluding the dynamical contribution of planetary companions.}. 
Each DWD is expected to merge in 150 Myr and 430 Myr (Tab \ref{tab:merge}). 

Another way to test the possibility of planetary system detection by LISA is to check the acceleration of the center of mass (CoM). When the acceleration is non negligible, it will leave an imprint on the GW signal; an apparent acceleration along the line of sight will cause each DWD to exhibit an apparent GW frequency chirp \citep[e.g.,][]{Xuan2023,Ebadi2024}. 
To qualitatively illustrate this point, we selected two systems from Population B (\#126 and \#235, Tab. \ref{tab:merge}) with the largest CoM shift, and we measured the average acceleration over a 8 years timescale (i.e., the duration of LISA extended mission). For both systems we see an increase of the acceleration as a consequence of planetary ejections from the system (see Fig. \ref{fig:COM_acceleration} for system ID \#235; the CoM acceleration of system ID \#126 reaches a similar average acceleration but earlier in time). Both systems had an average acceleration $a_{\rm CoM}$ of $\sim$ 0.4 au/yr$^2$, followed by an average acceleration of 0.9 au/yr$^2$ after ejections happened. Furthermore, the dispersion of the acceleration also increased, passing from $\Delta a_{\rm CoM}$ = 0.6 au/yr$^2$ to $\Delta a_{\rm CoM}$ = 1.0 au/yr$^2$. In terms of timescales, ejections began at 2.00$\times$10$^{3}$ yr and 9.22$\times$10$^{5}$ yr, and occurred over a time-frame of 1.80$\times$10$^{4}$ yr and 1.99$\times$10$^{5}$ yr for system ID \#126 and ID \#235, respectively.
At the time of writing the minimum $a_{\rm CoM}$ for a system to be detected by LISA has not been defined yet. Future studies focusing on the planetary induced accelerations are to be developed, and will be able to address whether the acceleration of the CoM of the systems presented here is sufficient or not to improve the planetary detection rates of LISA. 

\subsection{Caveats and model limitations}
Before drawing our conclusions, we want to emphasize that, due to the computational cost of the simulations, our integration times probe the early phases of dynamical evolution of DWD systems, whose lifetime is significantly longer than what modeled in this study. Furthermore, we stress that stability studies are in general sensitive to the assumed initial conditions, therefore our results are valid within the limits of the assumed priors and sampled parameter space. In terms of physical assumptions, we note that our simulations do not account for general relativity effects, tidal interaction, binary period shrinkage, secular resonances and radiation forces, as the addition of such physical processes is beyond of the scope of the current work. Finally, the discussion on the prospective  detectability of similar systems with LISA should be regarded as  illustrative, since it relies on possible synthetic architectures of second-generation multi-planet systems. 

\section{Conclusions}
\label{sec:conclusions} 
This work has been developed within the framework of the planetary detection science case of the Laser Interferometer Space Antenna mission, which will greatly advance our understanding of the universe via gravitational wave astronomy. The mission will shed light on a variety of unanswered questions, and one of these is whether planets can form and/or survive around binary systems where both stars evolved into white dwarfs. 
Although circumbinary planets have been found around evolved systems, to date none of these is orbiting a DWD. The most probable reason behind the lack of detection is the selection bias in the observations due to the faintness of the stellar hosts. 
\cite{kostov} and more recently \cite{Columba2023} proved that planets can survive the catastrophic events that take place during the evolution of a binary system. At the same time \cite{Ledda2023} showed that a second generation of giant planets can form from the ejected material that remains bound to the binary. The research work we performed takes us a step ahead: with N-body simulations we investigate the orbital evolution of giant planets around DWDs to see whether the planets that formed within a second-generation circumbinary disc are dynamically stable and can survive for us to observe them, or get destabilized.

Our study shows that, in the event of giant planet formation around DWDs, these planets can survive over multiple million years after disc dispersal (see Tables \ref{tabEventiPopAB}, \ref{tab.ejectedbodAB} and \ref{tabeventiNAMDmagAB} for a summary of the results). In the sampled parameter space, we find that more massive binaries hosting more massive planets, our Population B and C, are more likely to go through unstable phases and experience catastrophic events compared to the less massive counterpart, our Population A (Figures \ref{heatmapA}, \ref{heatmapDWD3} and \ref{heatmapC}). The most common catastrophic event that our systems experience is the scattering/ejection of planets. In the most extreme scenarios, where multiple catastrophic events take place, and especially when planets experience close encounters with the binary, the planetary systems are disrupted. Considering all the 2500 simulated planetary systems, 7.7\% are disrupted while 92.3\% have at least one planet that survives the dynamical evolution. Dynamically hot systems, which are the product of chaotic evolution, always evolve towards less compact architectures and the most massive planet contains larger fractions of the final planetary mass. In addition, dynamically hot systems can experience large shifts in the center of mass, even up to $\sim$30 au (Figure \ref{SsvsCOMDWD4_resonant} and bottom panels of Figures \ref{SsvsSmCOM_A}, \ref{SsvsSmCOMDWD3} and \ref{SsvsSmCOMDWD3_resonant}).
On the contrary, dynamically cold systems, which experience an ordered evolution, preserve their initial orbital architecture and do not experience mass loss. 

Comparing the non-resonant with the initially resonant configurations, our results indicate that planetary systems born in resonance tend to preserve dynamically cold configurations over time. While resonant setups overall enhance the stability of low-multiplicity systems (two or three planets), higher-multiplicity resonant systems prove instead more prone to instability and planet loss. In particular, nearly all four-planet systems and all five-planet systems starting in resonant chains ultimately break the resonance by losing at least one planet.

Looking at all four populations as a whole, we find that systems with higher initial multiplicity are more likely to undergo unstable phases, where ejections and collisions can take place. We also find that the final planetary multiplicity is anti-correlated with their dynamical excitation (recorded by their NAMD value), in agreement with the findings of \cite{turrini2020} for observed multi-planetary systems (Figure \ref{NAMD-multiplicity}). Two-planet systems remain dynamically cold in the majority of cases and preserve all their original planets. On the other hand, the five-planet population proved highly prone to disruption, while the four-planet population decreased by 56.1\% and the three-planet population decreased by 22.5\%. These decreases led to an increase (122\%) of the two-planet population and the creation of a new single-planet population (7.1\% of the totality of the systems). Our findings are consistent with the interpretation of \cite{turrini2020}, suggesting that low-multiplicity systems are often the outcome of dynamical instabilities that removed some of their primordial planets.

Furthermore, concerning the single-planet systems, most of them, given their planet mass and period, have the potential to be detected by LISA, provided that their distance, sky location, polarization and inclination allow for a sufficient GW signal-to-noise ratio for the systems to be detected. 
In terms of GW detection of multi-planet systems, we provide a formula (Equation \ref{deltaF}) to approximate  the global frequency shift that the planets would exert onto the GW produced by the DWD. For the systems studied here, this frequency shift is for the majority of the cases smaller than LISA frequency resolution, indicating that these multi-planet systems are unlikely to be detectable by LISA. For a handful of resonant systems (Pop C) the frequency shift caused by planets in the same system, is large enough to be detected within 8 years of LISA observations. 
Nevertheless,  specific Bayesian studies accounting for both multi-body waveform and the orbital shrinking of the binary, should to be performed to confirm/refute the detectability (such as the work done by \citealt{Katz2022} on single-planet systems). 
It is important to approach these detectability considerations as illustrative of LISA's potential capabilities for planetary detection, and not as predictions, particularly since they are based on synthetic architectures.
Finally, we have studied the acceleration of the center of mass of two systems which presented the largest inward and outward CoM shift during their evolution. An acceleration of the CoM can leave an imprint on the GW signal which can improve the system detectability. 
So far no acceleration threshold value has been studied for the LISA planetary system case, yet our two systems have an average acceleration of 0.9 au/yr$^2$. If the LISA threshold value would prove to be smaller than 0.9 au/yr$^2$, $Magrathea$ planets would have one more chance to be found by gravitational wave astronomy. 

\begin{acknowledgements}
We thank N. Tamanini for the helpful discussions. This work has been developed within the framework of the ``LISA Triple and CBPs'' project of the LISA Astrophysical WG of the ESA LISA mission Consortium.
A.N. acknowledges support from the Swiss National Science Foundation (SNSF) under grant PZ00P2\_208945.
D.T. and D.P.  acknowledge support by the Fondazione ICSC, Spoke 3 ``Astrophysics and Cosmos Observations'', National Recovery and Resilience Plan (Piano Nazionale di Ripresa e Resilienza, PNRR) Project ID CN\_00000013 ``Italian Research Center on High-Performance Computing, Big Data and Quantum Computing'' funded by MUR Missione 4 Componente 2 Investimento 1.4: Potenziamento strutture di ricerca e creazione di “campioni nazionali di R\&S (M4C2-19)'' - Next Generation EU (NGEU). D.T. and D.P. also acknowledge support from the ASI-INAF grant no. 2021-5-HH.0 plus addenda no. 2021-5-HH.1-2022 and 2021-5-HH.2-2024, the COST Action CA22133 PLANETS.
D.T. also acknowledges the European Research Council via the Horizon 2020 Framework Programme ERC Synergy ``ECOGAL'' Project GA-855130. 
 C.D. acknowledges financial support from the grant RYC2023-044903-I funded by MCIU/AEI/10.13039/501100011033 and by
the ESF+, and from the INAF initiative ``IAF Astronomy Fellowships in Italy'', grant name \textit{GExoLife}.
\end{acknowledgements}

\bibliographystyle{aa} 
\bibliography{Biblio} 

\end{document}